\documentstyle[a4,epsfig,12pt]{article}
\newcommand{\beq}{\begin{equation}}
\newcommand{\eeq}{\end{equation}}
\begin{document}
\title{Space-time description of the hadron interaction at
high energies.}
\author{V.N.Gribov}

\date{}

\maketitle

\begin{abstract}
In this lecture we consider the strong and electromagnetic interactions
of hadrons in a unified way.
It is assumed that there exist point-like particles ({\em partons}) in
the sense of quantum field theory and that a hadron with large momentum
$p$ consists of $\sim\ln(p/\mu)$ partons
which have restricted
transverse
momenta, and longitudinal momenta
which range from $p$ to zero.
The density of partons increases with the
increase of the
 coupling constant. Since the
probability of their recombination also increases,
an equilibrium may be reached.
In the lecture we will consider consequences of the
  hypothesis that the
 equilibrium really {\em occurs}.
We  demonstrate that it
leads to constant total cross
sections at high energies, and to
the  Bjorken scaling
in the deep inelastic ep-scattering.
The asymptotic value of the total cross sections
of hadron-hadron scattering turns out to be
universal,
while the cross sections of quasi-elastic
scattering processes at zero
angle  tend to zero.

The multiplicity of the outgoing hadrons and their
distributions
in longitudinal momenta (rapidities) are also
discussed.
\end{abstract}

\section*{Introduction}

In this lecture we will try to describe electromagnetic
and strong interactions of hadrons in the same framework which follows
from general quantum field theory considerations without the introduction
of quarks or other exotic objects.

We will assume that there exist
 point-like constituents in the sense of quantum field
theory which are, however, strongly interacting. It is convenient to refer
to these particles as {\em partons}. We will not be interested
in the quantum numbers of these partons, or the symmetry
properties of their interactions.
We will assume that, contrary to
the perturbation theory, the integrals over the transverse momenta of
virtual particles converge like in
the $\lambda \varphi^3$ theory. It turns
out that within this picture a common
cause exists for two seemingly
 very different phenomena:
the Bjorken scaling in deep inelastic
scattering, and the recent theoretical
observation that all hadronic cross sections
 should approach the  same limit
(provided that the Pomeranchuk pole exists).
The lecture is organized as follows. In
the first part we discuss the
propagation of the hadrons in the space as a process of creation and
absorption of the virtual particles (partons) and formulate the notion
of the parton wave function of the hadron. The second part
describes momentum and coordinate parton distributions
in hadrons. In the third part we consider the process of
 deep inelastic scattering.
It is shown that from the point of view of our approach the deep
inelastic scattering
satisfies the Bjorken scaling,
 and, in contrast to   the quark model,
the multiplicity of the produced hadrons is of the order of
$\ln \frac{\nu}{\sqrt{q^2}}$. The fourth part is devoted to the strong
interactions of hadrons
and it is shown that in the same framework
the total hadron cross sections have to approach asymptotically
the same limiting value. In the last part of the lecture we discuss the
processes of elastic and quasi-elastic scattering at high energies. It
is demonstrated that the cross sections of the quasi-elastic scattering
processes at zero angle tend to zero
at asymptotically high energies.

Let us discuss, how one can think of the space-time propagation
of a physical particle in terms of virtual particles which are involved
in the interaction with photon and other hadrons.
It is well known that the propagation of a real particle
is described by its Green function, which corresponds to a series of
Feynman diagrams of the type
\begin{figure}[ht]
\begin{center}
\epsfxsize=14cm
\epsfbox{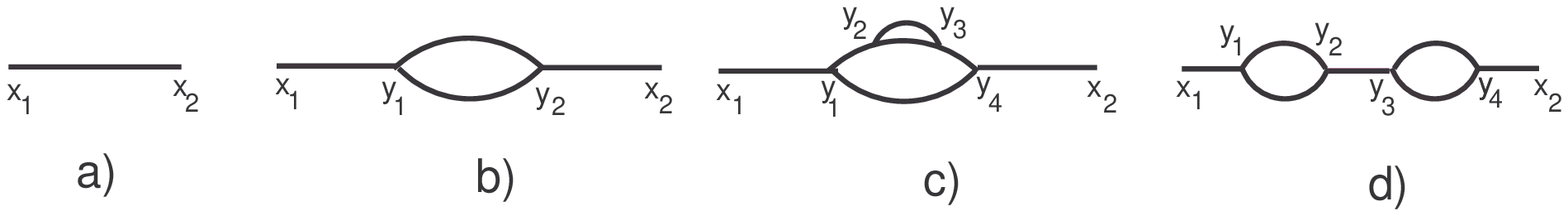}
\end{center}
\caption{}
\end{figure}
(for simplicity, we will consider identical scalar particles). The
Feynman diagrams, having many remarkable properties, have, nevertheless,
a disadvantage compared to the old-fashioned perturbation theory.
Indeed, they do not show how a system evolves with time in a given
coordinate reference frame.
For example, depending on the relations between the time coordinates
$x_{10}$, $y_{10}$, $x_{20}$ and $y_{20}$, the graph in Fig.1b
corresponds to different processes:
\begin{figure}[ht]
\begin{center}
\begin{picture}(100,100)
\put(-120,-20){
\epsfxsize=14cm
\epsfbox{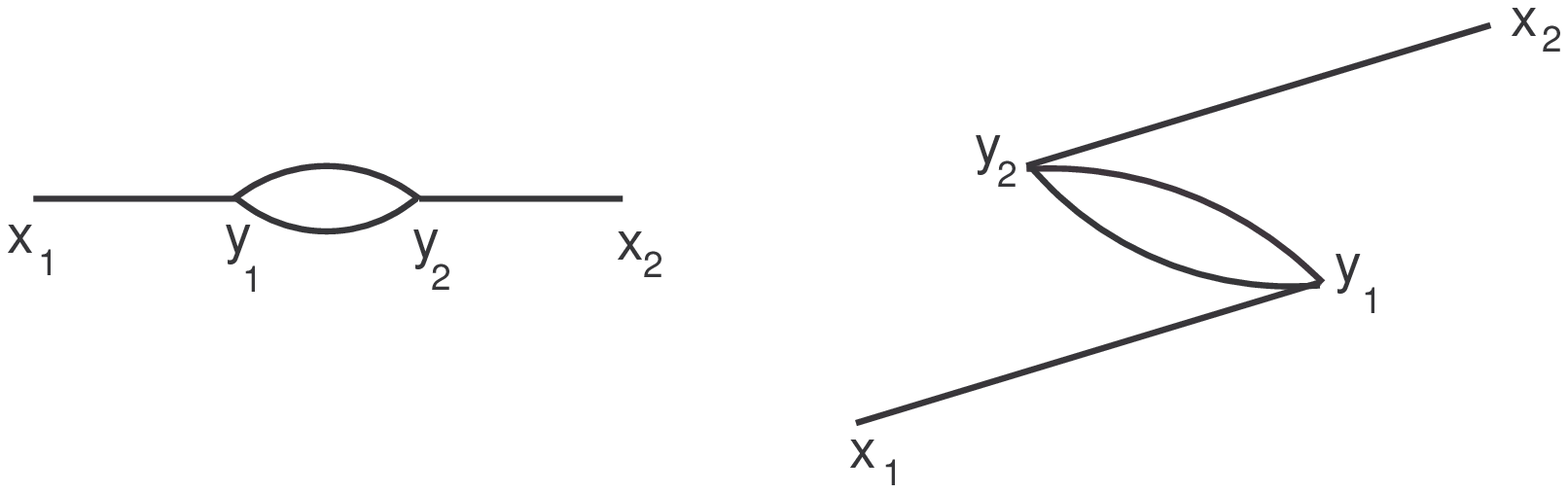}
}
\end{picture}
\end{center}
\caption{}
\end{figure}

Similarly, the diagram Fig.1c corresponds to the processes

\newpage
\begin{figure}[ht]
\begin{center}
\begin{picture}(100,100)
\put(-150,-20){
\epsfxsize=14cm
 \epsfbox{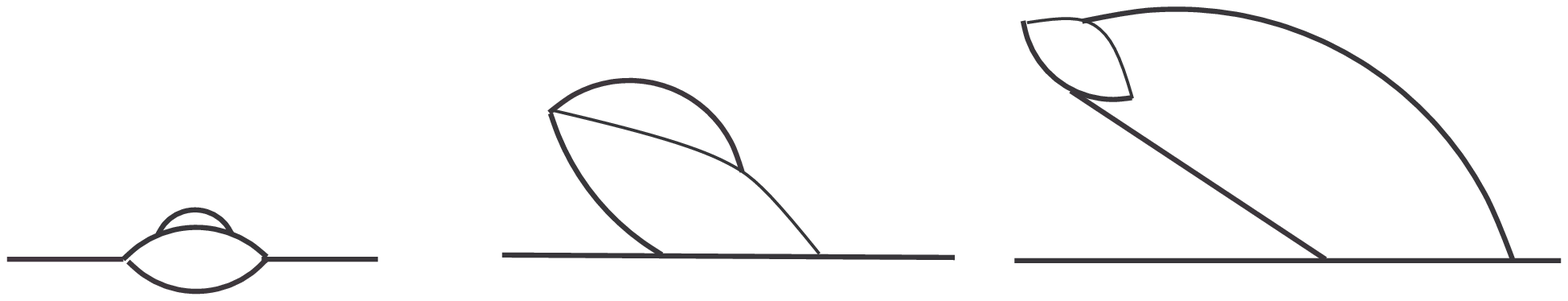}
}
\end{picture}
\end{center}
\caption{}
\end{figure}

In quantum electrodynamics, where explicit calculations can be carried
out, this complicated correspondence is of little interest. However,
for strong interactions, where explicit calculations are impossible,
distinguishing  between different space developments will be useful.

Obviously, if the interaction is strong (the coupling constant $\lambda$
is large), many diagrams are relevant.
The first question which arises is which
configurations dominate: the ones
which correspond to the subsequent decays
of the particles -  the diagrams
Fig.2a and Fig3.a, or those which
 correspond to the interaction of the
initial particle with virtual "pairs" created in the vacuum. It is clear
that if the coupling constant is large and the
 momentum of the incoming
particle is small (see below),
 configurations with "pairs" dominate
(at least if the theory does not contain infinities). Indeed, if $x_{20}
- x_{10}$ is small, then
in the case of configurations without "pairs" the
integration regions corresponding
 to each correction will tend to zero
with an increase of the number of corrections. At the same time, for the
configurations containing "pairs" the region of integration over time
will remain infinite. Hence, if the retarded Green function
$G^r(y_2-y_1)$ does not have a strong singularity at $x_{20}-x_{10}
\rightarrow 0$,
the contribution of the configurations
without  "pairs" will be
relatively small if the coupling constant is large. Even the graphs of the
type Fig.1d are determined mainly by configurations like
\begin{figure}[ht]
\begin{center}
\begin{picture}(200,200)
\put(30,0){
\epsfxsize=6cm
\epsfbox{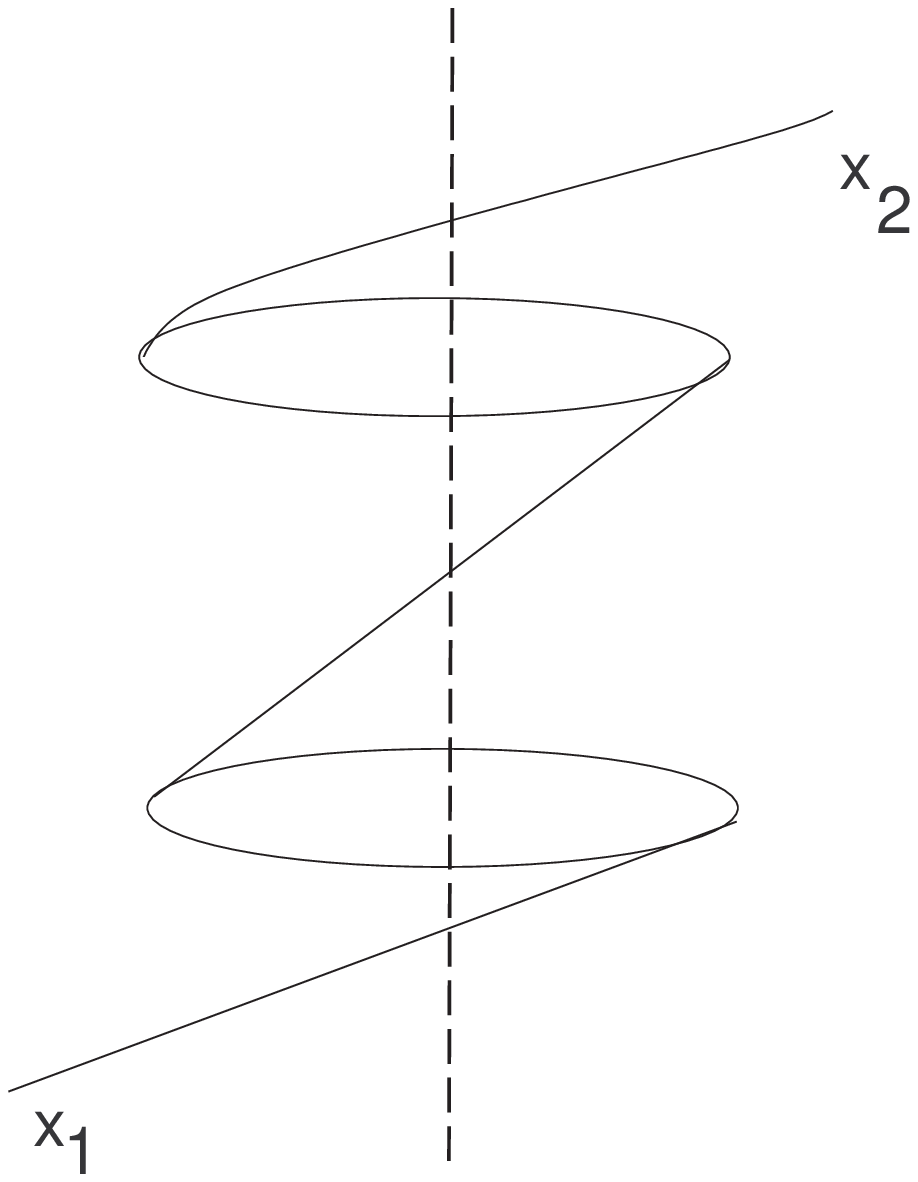}
}
\end{picture}
\end{center}
\caption{}
\end{figure}

This means that if we observe a low energy particle at any particular
moment of time (the cut in the diagram in Fig. 4), we will see few partons
which are decay products of the particle, and a large number of
virtual "pairs"
which will interact with these partons in the future.

What happens if a particle has a large momentum in
our coordinate reference frame? To analyze the space-time evolution of a
fast particle  we have to consider
a space-time interval $(x_1-x_2)^2$ such
that $(x_1-x_2)^2\sim\frac{1}{\mu^2}$, and
$t_2-t_1\sim E/\mu^2\gg 1/\mu$. Here $\mu$ is the mass of the particle,
$E$ its energy.
In this case, $\vec{x}_1-\vec{x}_2 = \vec{v}(t_2-t_1)$,
$(x_2-x_1)^2=\frac{\mu^2}{E^2}(t_2-t_1)^2 \sim \frac{1}{\mu^2}$. For such
intervals the relation between
 the configurations with and without "pairs" changes.
Configurations corresponding to a decay of
one parton into many others start to dominate,
 while the role of configurations with "pairs" decreases.

The physical origin of
this phenomenon is evident.
A fast parton can decay, for
example, into two fast partons which, due to the energy-time
uncertainty relation,
will exist for a long time (of the order of $E/\mu^2$), since
\[
\Delta E = \sqrt{\mu^2+\vec{p}^2} - \sqrt{\mu^2+\vec{p}_1^2} -
 \sqrt{\mu^2+(\vec{p}-\vec{p}_1)^2} \sim \frac{\mu^2}{2|\vec{p}|}-
 \frac{\mu^2}{2|\vec{p}_1|} - \frac{\mu^2}{2|\vec{p}-\vec{p}_1|}.
\]
Each of these two partons can again decay into two partons and this
will continue up to the point when slow particles, living for a time
of the order of $\frac{1}{\mu}$, are created. After that the
fluctuations must evolve in the
 reverse direction, i.e. the recombination
of the particles begins.

On the other hand, due to the same uncertainty relation,   creation
of virtual "pairs" with large momenta in vacuum
 is possible only for
short time intervals of the order of $\frac{1}{p}$. Hence, it
affects only the region of small momentum partons.
The way in which this phenomenon manifests itself can be seen using
the simplest graph in Fig.5. as an example. We will observe that it is
possible to place here many
emissions  in spite of the fact that the interval $x_{12}^2$ is of the
order of unity ($1/\mu^2$), and the Green function depends only on the
invariants.
\begin{figure}[h]
\begin{center}
\begin{picture}(200,200)
\put(-80,0){
\epsfxsize=14cm
\epsfbox{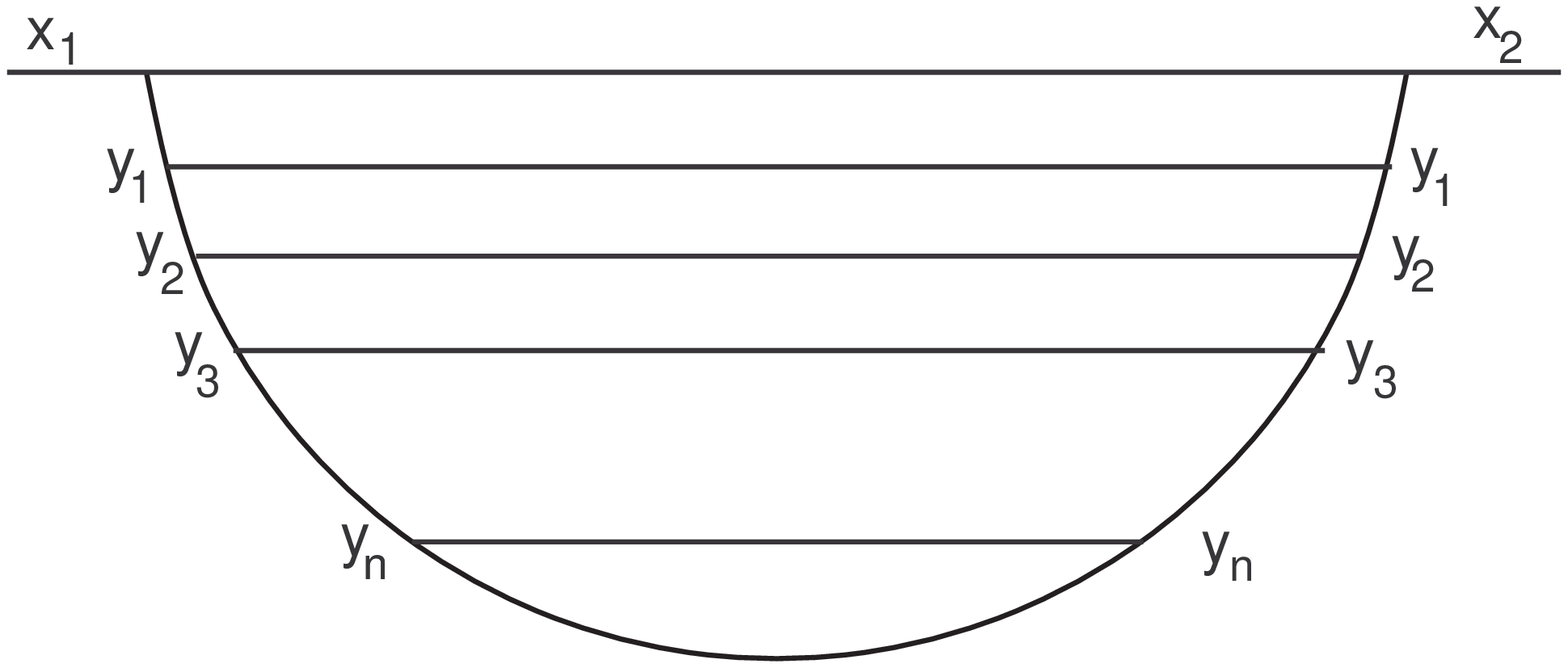}
}
\end{picture}
\end{center}
\caption{}
\end{figure}

For the sake of simplicity, let us verify this for one space
dimension ($y_i=(t_i,z_i)$). Suppose that $x_1=(-t,-z)$ and $x_2=(t,z)$,
$x_{12}^2=(2t)^2-(2z)^2$. Then $t=z+\frac{x_{12}^2}{8z}$. Let us choose
the variables $y_i$, $y'_i$ in the same way: $y_i=(-t_i,-z_i)$,
$y'_i=(t'_i,z'_i)$, and consider the following region of integration in
the integral, corresponding to the diagram in Fig.5:
\[
1<z_n<z_{n-1} \ldots <z_1<t,
\]
\[ 1<z'_n<z'_{n-1} \ldots <z'_1<t,
\]
\[ z_i \sim z'_i, \quad t_i=z_i+\frac{y_i^2}{2z_i}, \quad
 t'_i=z'_i+\frac{y^{'2}_i}{2z'_i} .
\]
The integrations over $d^2 y_1 \ldots d^2 y_n d^2 y'_1 \ldots d^2 y'_n$
can be substituted by integrations over $y_i^2$, ${y'_i}^2$
and $z_i\equiv y_{iz},z_i^{\prime}\equiv y^\prime_{iz}$
\[
d^2y_i=\frac{1}{2}dy_i^2\frac{dz_i}{z_i}, \qquad
  d^2y'_i=\frac{1}{2}dy^{'2}_i\frac{dz'_i}{z'_i}
\]
It is easy to see that in this region
 of integration the arguments of
all Green functions:
$(y_i-y'_i)^2$, $(y_i-y'_{i+1})^2$,
$(y'_i-y'_{i+1})^2$ ,
are of the order of unity, and the integrals do
not contain any small factors. All these conditions for $y_i$ can
be satisfied simultaneously for a large number of
emissions:
$n\sim \ln t$. Indeed, if we write $z_n$ in the form $z_n=C^n$, all
conditions will be fulfilled for
\[
n \sim \frac{\ln t}{\ln C} , \qquad C \geq 1.
\]
Obviously, one can consider a more complicated diagram than Fig.5
by including interactions of the virtual particles. On the other hand, configurations
containing vacuum "pairs" play a minor role. Moving backwards in time
is possible only for short time intervals (Fig.6):
\begin{figure}[h]
\begin{center}
\begin{picture}(100,100)
\put(-120,0){
\epsfxsize=14cm
\epsfbox{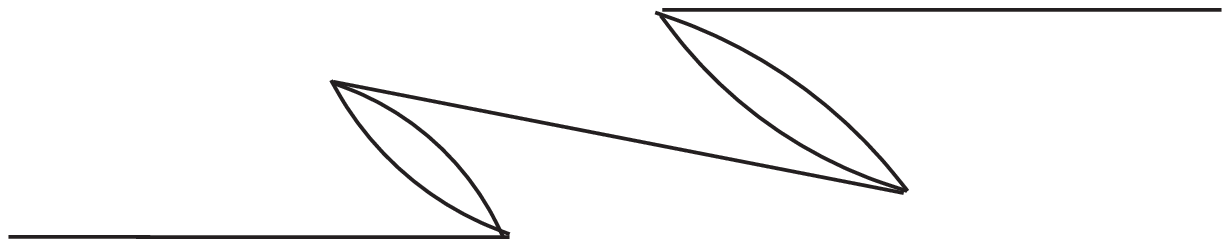}
}
\end{picture}
\end{center}
\caption{}
\end{figure}

\vspace*{0.5cm}
Hence, we reach the following picture. A real particle with a large
momentum $p$ can be described as an ensemble of an indefinite number
of partons of the order of $\ln\frac{p}{\mu}$ with momenta
in the range from $p$ to zero, and several vacuum pairs with small momenta
which in the future can interact with the target.

The observation of a slow particle during an interval of the order of
$1/\mu$ does not tell us anything about the structure of the particle
since we cannot distinguish it from the background of the vacuum
fluctuations, and we can speak only about the interaction of particles or
about the spectrum of states. On the contrary, in the case of a fast
particle we can speak about its structure, i.e. about the fast partons
which do not mix with the vacuum fluctuations. As a result, in a certain
sense a fast particle becomes an isolated system which is only weakly
coupled to the vacuum fluctuations. Hence, it can be described using a
quantum mechanical wave function or an ensemble of wave functions, which
determine probabilities of finding certain numbers of partons and their
momentum distribution. Such a description is not invariant , since the
number of partons depends on the momentum of the particle, but it can
be considered as covariant.
Moreover, it may be even invariant, if the momentum distribution of the
partons is homogeneous in the region of momenta
much smaller  than the
maximal one, and much larger than $\mu$.

Indeed, under the
transformation  from one reference frame to another in which
the particle has, for example, a larger momentum, a new region
emerges in the distribution of partons;  in the old region, however,
 the  parton distribution remains unchanged. One usually describes
hadrons in terms of the quantum mechanics of partons in
the  reference frame
which moves with an infinite momentum, because in this case all partons
corresponding to vacuum fluctuations have zero momenta, and such a
description is exact. Such a reference frame is convenient for the
description of the deep inelastic scattering. However, it is not as
good for describing strong interactions, where the slow partons
are important. In any case, it appears useful to preserve the
freedom in choosing the reference frame and to use the
covariant description. This allows a more effective analysis of the
accuracy of the derivations.

\section{Wave function of the hadron. Orthogonality and normalization}

The previous considerations allow us to introduce the hadron wave
function in the following way. Let us assume, as usual, that at
$t\longrightarrow -\infty$ the hadron can be represented as a bare particle
(the parton). After a sufficiently long time the parton will decay
into other partons and form a stationary state which we call a hadron.
Diagrams corresponding to this process are shown in Fig.7.
\begin{figure}[ht]
\begin{center}
\begin{picture}(250,150)
\put(-50,-40){
\epsfxsize=12.0cm
\epsfbox{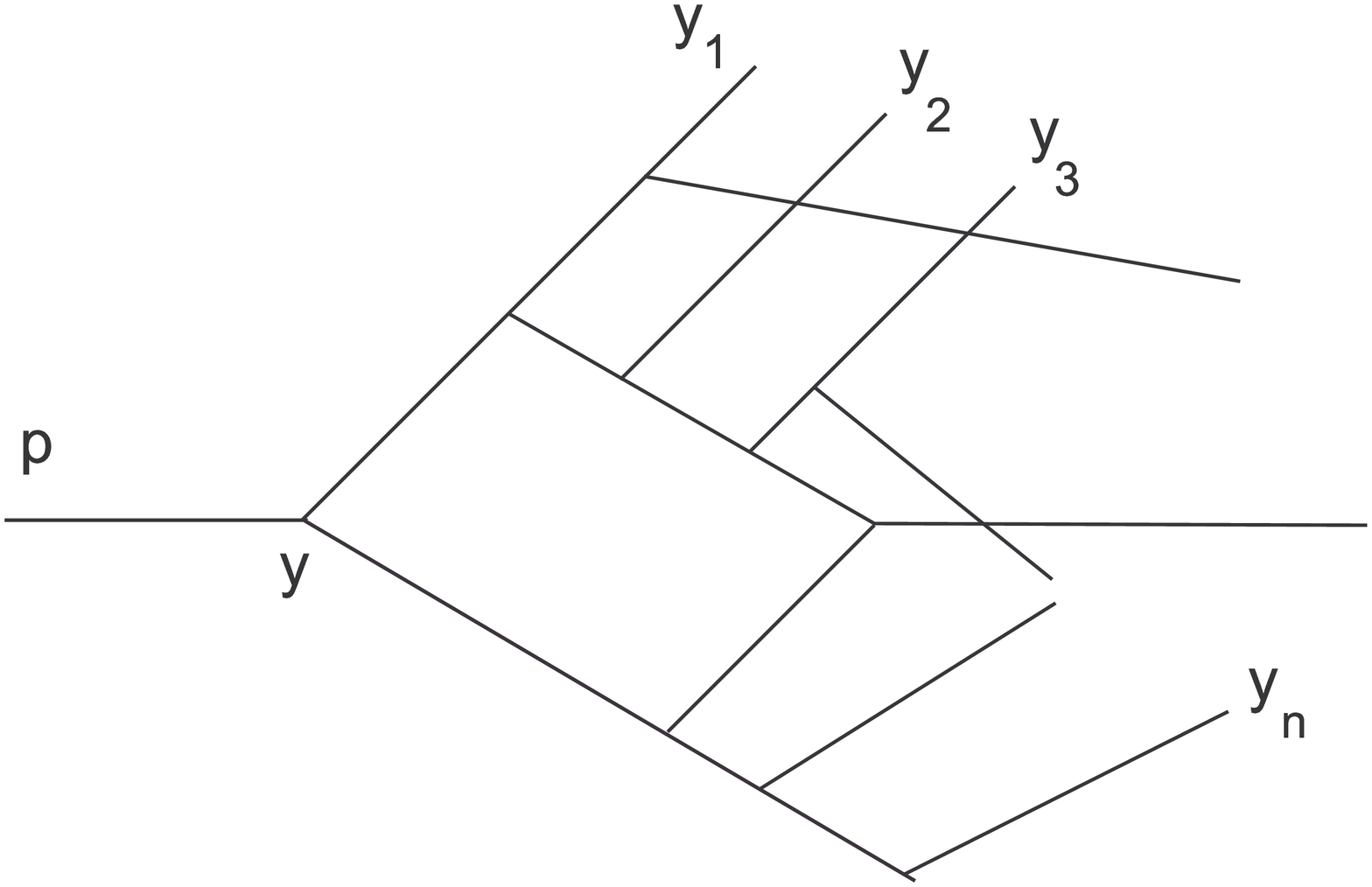}
}
\end{picture}
\end{center}
\caption{}
\end{figure}

Let us exclude from the Feynman diagrams those configurations (in the
sense of integrations over intermediate times) which correspond to
vacuum pair creation.


For the theory $\lambda\varphi^3$ such a separation of vacuum fluctuations
corresponds to decomposing  $\varphi$ into positive and negative
frequency parts $\varphi = \varphi^+ + \varphi^-$ and substituting
$\varphi^3 = (\varphi^+ + \varphi^-)^3$ by $3(\varphi^{-2}\varphi^+ +
\varphi^-\varphi^{+2})$. The previous discussion shows that the ignored
term $\varphi^{+3} + \varphi^{-3}$ would mix only partons with small
momenta.

It is natural to consider the set of all possible diagrams with a given
number of partons $n$ at the given moment of time as a component of
the hadron wave function $\Psi_n(t,\vec{y}_1,\ldots,\vec{y}_n,p)$.
Similarly, we can determine the wave functions of several hadrons
with large momenta provided the energy of their relative
motion is small compared to their momenta. The
latter condition is necessary to ensure that
slow partons are not important in  the interaction. The Lagrangian
of the interaction remains Hermitian even after
the terms corresponding to
the vacuum fluctuations are omitted.
As a result, the wave functions will be orthogonal,
and will be normalized in the usual way:
\begin{equation}
\label{1}
\sum_n\int \Psi^{b^*}_n(\vec{y}_1\ldots,\vec{y}_n,p_b)i
\stackrel{\leftrightarrow}{\partial}
\Psi^{a}_n(\vec{y}_1\ldots,\vec{y}_n,p_a)
\frac{d^3y_1\ldots d^3y_n}{n!}=(2\pi)^3\delta(\vec{p}_a-\vec{p}_b)
\delta_{ab},
\end{equation}
or similarly in the momentum space, after separating
\begin{equation}
\label{2}
\sum_n\frac{1}{n!} \int \Psi^{b^*}_n(\vec{k}_1\ldots,\vec{k}_n,\vec{p})
\Psi^{a}_n(\vec{k}_1\ldots,\vec{k}_n,\vec{p})
\frac{d^3k_1\ldots d^3k_n}{2k_{10}\ldots 2k_{n0}}\frac{\delta(p-\sum k_i)}{(2\pi)^{3n-1}}
=\delta_{ab}.
\end{equation}
For the momentum range $k_i\gg \mu$, the wave
functions coincide with those calculated in the infinite momentum frame.
In this reference frame they do not depend on the momentum of the system
(except for a trivial factor). This can be easily proven by expanding
the parton momenta
\begin{equation}
\label{3}
\vec{k}_i=\beta_i\vec{p}+k_{i\perp},
\end{equation}
and writing the parton energy in the form
\begin{equation}
\label{4}
\varepsilon_i=\sqrt{\vec{k}_i^2+m^2}=\beta_i p+\frac{m^2+k_{i\perp}^2}
{2p\beta_i}.
\end{equation}
Note now that the integrals which determine $\Psi_n$, corresponding
to  Fig.7, can be represented in the form of the old-fashioned
perturbation theory where only the differences between the energies of
the intermediate states and the initial state $E_k-E$
enter, and the momentum is conserved.
Hence, the terms linear in $p$ cancel in these differences, and
concequently
\beq
\label{5}
E_k-E=\frac{1}{2p}\left( \sum_i\frac{m^2+k_{i\perp}^2}{\beta_i}-m^2\right)
\eeq
Each consequent intermediate state in Fig.7 in the $\lambda\varphi^3$
model differs from the previous one by the appearance or disappearance
of one particle. The
factor $\frac{1}{k_i}=\frac{1}{2p}\frac{1}{\beta_i}$, which comes from the
propagator of this particle, cancels $2p$ in (\ref{5}). Hence, there
remain only integrals over $d^2 k_{i\perp}\frac{d\beta_i}{\beta_i}$, and the
resulting expression does not depend on $p$.
\begin{equation}
\label{6}
\sum_n\frac{1}{n!}\int \Psi^{b^*}_n(k_{i\perp},\beta_i)
\Psi^{a}_n(k_{i\perp},\beta_i)\prod\frac{d^2k_{i\perp}}{2(2\pi)^2}
\frac{d\beta_i}{\beta_i} (2\pi)^3\delta(1-\sum\beta_i)=\delta_{ab} .
\end{equation}
For slow partons,
where the expansion (\ref{4}) is not correct, the
 dependence on momentum $p$ does not disappear, and contrary to
 the case of the system moving with $p=\infty$,
this dependence cuts off the sum over the number of partons.

\section{Distribution of the partons in space and momentum}

The distribution of partons in longitudinal momenta can be characterized
by the rapidity:
\begin{equation}
\label{7}
\eta_i=\frac{1}{2}\ln\frac{\varepsilon_i+k_{iz}}{\varepsilon_i-k_{iz}},
\end{equation}
where $k_{iz}$ is the component of the parton momentum along the
hadron momentum.
\begin{equation}
\label{8}
\eta_i\approx\ln\frac{2\beta_i p}{\sqrt{m^2+k^2_{i\perp}}}.
\end{equation}
As it is well known, this quantity is convenient since it simply transforms
under the Lorentz transformations along  the  $z$
 direction:
$\eta_i^\prime=\eta_i+\eta_0$ , where
$\eta_0$ is the rapidity of the coordinate system.

The determination of
 the  parton distribution over $\eta$ is based on the
observation that in each decay process $k_1 \rightarrow k_2 + k_3$
shown in Fig.7 the momenta $\vec{k}_2$ and $\vec{k}_3$ are, in the
average, of the same order. This means that in the process of subsequent
parton emission and absorption the rapidities of the partons change
by a factor of the order of unity. At the same time the overall
range of parton rapidities is large, of the order of $\ln\frac{2p}{m}$. This
implies that in the rapidity space we have short
range forces.

Let us consider the density of the distribution in rapidity
\begin{eqnarray}
\label{9}
\lefteqn{\varphi(\eta,k_\perp,p)=} \nonumber\\
 & & \sum_n\frac{1}{n!}\int \left|
\Psi(k_\perp,\eta,k_{\perp 1},\eta_1,\ldots,k_{\perp n},\eta_n,)
\right|^2(2\pi)^3\delta(\vec{p}-\vec{k}-\sum\vec{k}_i)\prod
\frac{dk_i d\eta_i}{2(2\pi)^3}
\end{eqnarray}
in the interval $ 1 \ll \eta \ll \eta_p $ (see Fig.8).
\begin{figure}[ht]
\begin{center}
\begin{picture}(250,200)
\put(-50,-13){
\epsfxsize=12.0cm
\epsfbox{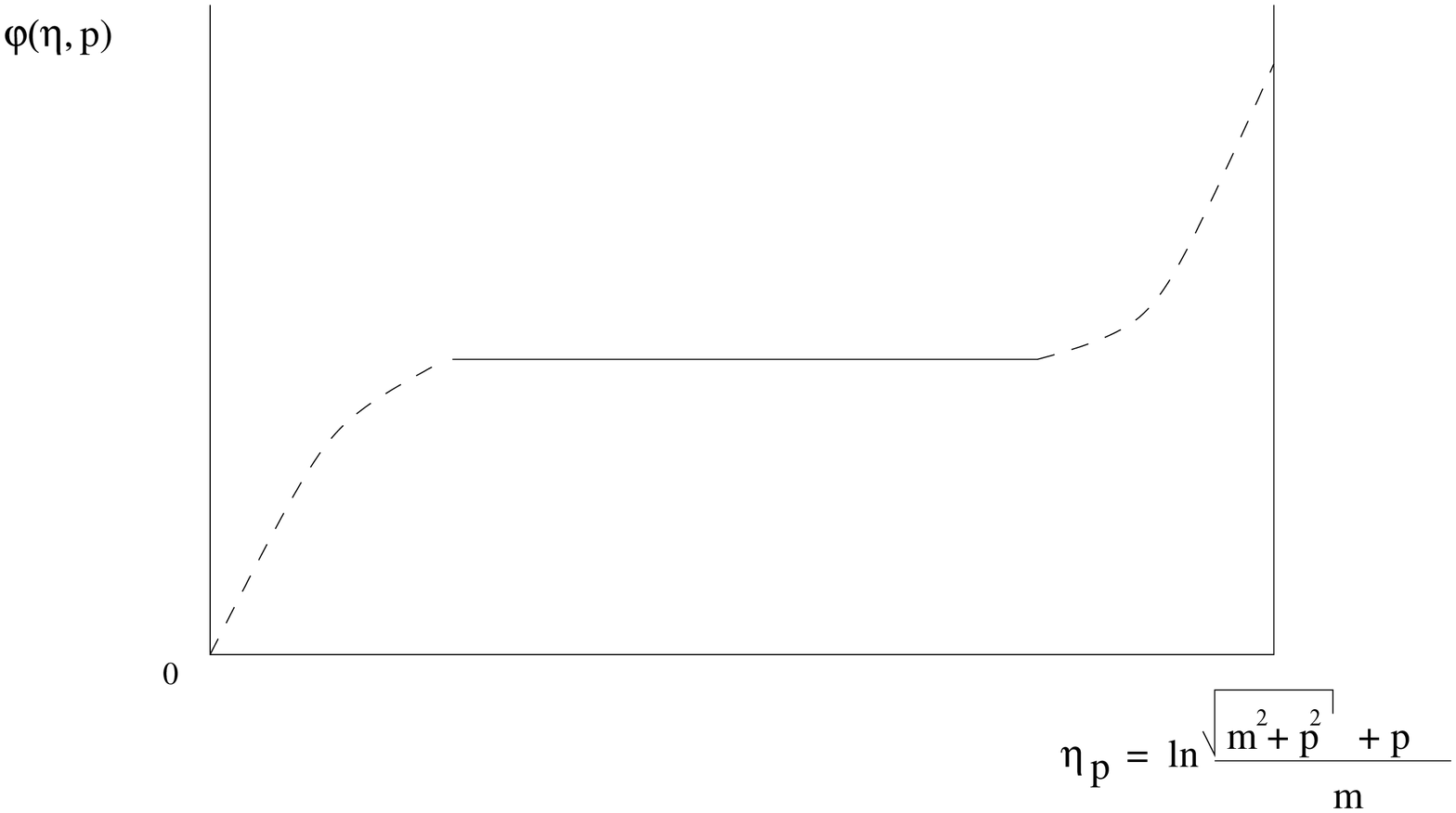}
}
\end{picture}
\end{center}
\caption{}
\end{figure}

The independence of $\varphi$ on $p$ for these
values of $\eta$  means that
$\Psi$ depends only on the differences $\eta_i-\eta_p$. If $\varphi =
\varphi(\eta-\eta_p, k_{\perp})$ decreases with the increase of
$\eta-\eta_p$, this corresponds to a weak coupling, i.e. to a small
probability of the decay of the initial parton. If the coupling constant
grows, the number of partons increases and at a {\em certain} value of
the coupling constant an equilibrium is reached, since the probability
of recombination also increases. The value of this critical coupling
constant has to be such that the recombination probability due to the
interaction should be larger than the recombination probability related
to the uncertainty principle.

The basic hypothesis is that such an equilibrium {\em does occur} and that
due to the short-range character of interaction it is {\em local}. This is
equivalent to the hypothesis of the constant total
cross sections of interaction at $p\rightarrow\infty$. Hence we assume that the
equilibrium is determined by the vicinity of the point $\eta$ of the order
of unity and it does not depend on $\eta_p$. Obviously, this can be
satisfied only if $\varphi(\eta,\eta_p,k_{\perp})=\varphi(k_{\perp})$
does not depend on $\eta$ and $\eta_p$ at $ 1 \ll \eta \ll \eta_p $.
According to the
idea of Feynman, this situation resembles
 the case of a sufficiently
long one-dimensional matter in which, due to the homogeneity of the
space, far from the boundaries the density is either constant or
oscillating (for a crystal). In our case the analogue of the homogeneity
of space is the relativistic invariance (the shift in the space of
rapidities). {\em For the time being} we will not consider the case
of the crystal. According to (\ref{9}), the integral of
$\varphi(\eta,\eta_p,k_{\perp})$ over $\eta$ and $k_{\perp}$ has the
meaning of the average parton density  which is, obviously, of
the order of $\eta_p \sim \ln\frac{2p}{m}$.

Generally speaking, we cannot say anything about the parton distribution
in the transverse momenta except for one statement: it is absolutely
crucial for the whole concept that it must be restricted to the region
of the order of parton masses, like in the $\lambda\varphi^3$ theory.

Consider now the spatial distribution of the partons.
First, let us discuss parton distribution in the plane
perpendicular to the momentum $\vec{p}$.
For that purpose it is convenient to transform from $\Psi_n(k_{1\perp},
\eta_1, k_{2\perp}, \eta_2,\ldots k_{n\perp}, \eta_n)$ to the
impact parameter representation
$\Psi_n(\vec{\rho}_1,\eta_1,\vec{\rho}_2,\eta_2,\ldots\vec{\rho}_n,\eta)$:
\begin{equation}
\label{10}
\Psi_n(\vec{\rho}_n,\eta_n)=\int e^{i\sum k_{i \perp}\rho_i}\Psi(k_{i\perp
},\eta_i)\delta(\sum k_{\perp i})(2\pi)^2\prod\frac{d^2k_i}{(2\pi)^2}.
\end{equation}
Let us rank the partons in the order of their decreasing rapidities.
Consider a parton with the rapidity $\eta \ll \eta_p$ and let us follow
its history from the initial parton.
Initially, we will assume
that it was produced solely
via parton emissions (Fig.9).
\begin{figure}[ht]
\begin{center}
\begin{picture}(250,280)
\put(-80,90){
\epsfxsize=6.0cm
\epsfbox{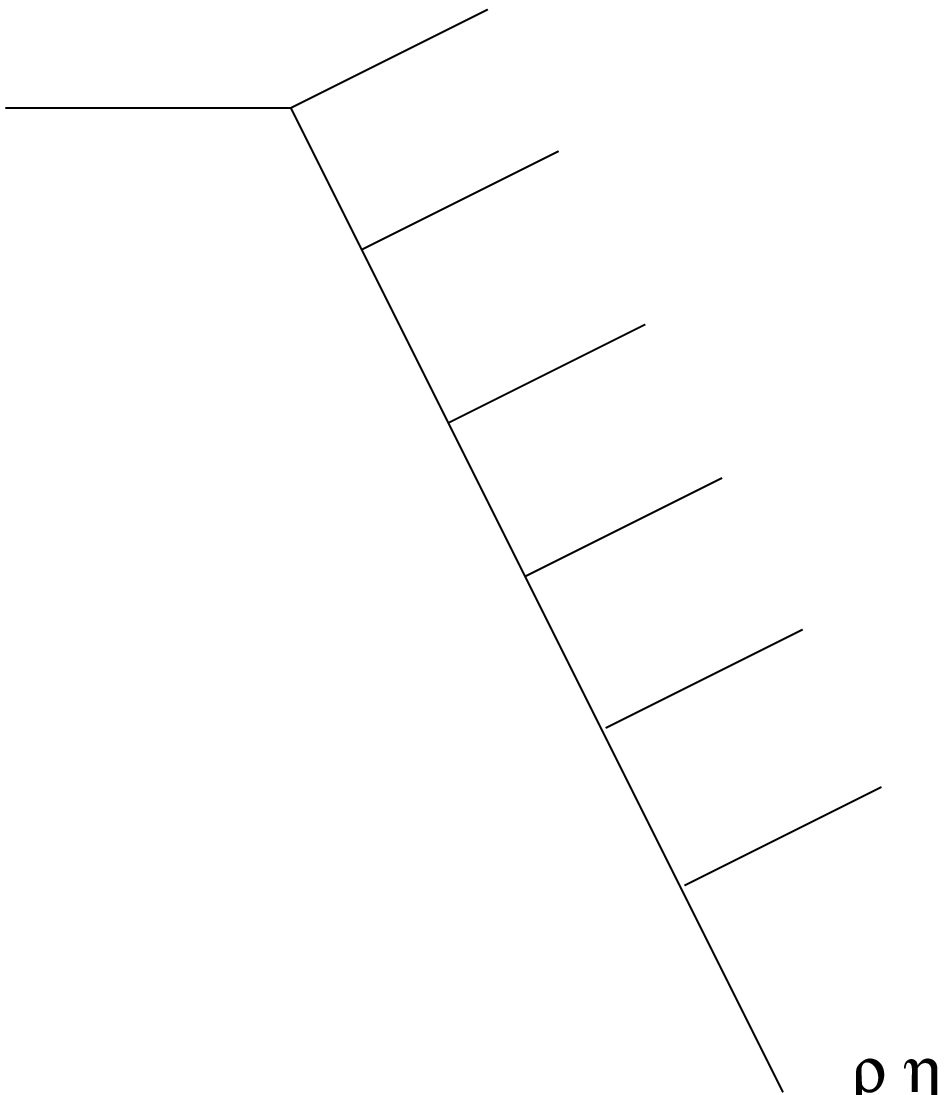}
}
\put(140,134){
\epsfxsize=4.5cm
\epsfbox{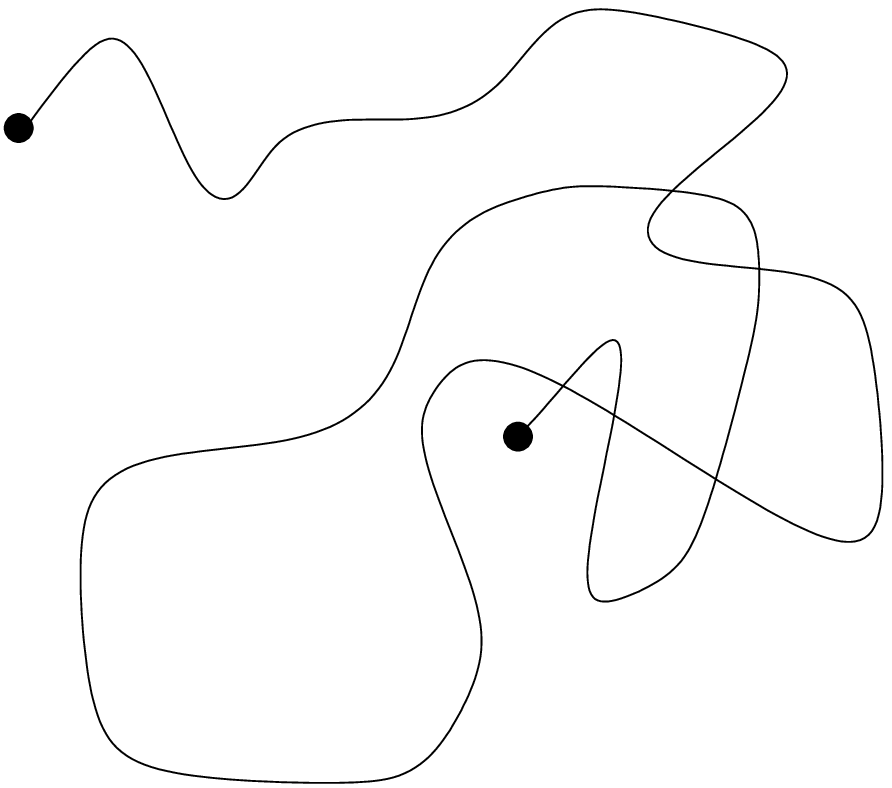}
}
\put(11,25){Figure 9. \hspace{3.8 cm} Figure 10.}
\end{picture}
\end{center}
\end{figure}
\setcounter{figure}{10}

In this case it is clear that if the transversal momenta of all partons
are of the order of $\mu$, than each parton emission leads to a change
of the impact parameter $\vec{\rho}$ by $\sim \frac{1}{\mu}$.
If $n$ emissions are necessary to reduce the rapidity from $\eta_p$
to $\eta$, and they are independent and random,
$\overline{(\Delta\rho)^2}\sim n$. If every emission changes the
rapidity of the parton by about one unit, then
\begin{equation}
\label{11}
\overline{(\Delta\rho)^2}=\gamma(\eta_p-\eta) .
\end{equation}
Hence, the process of the subsequent parton emissions results in
a kind of diffusion in  the impact parameter plane. The
parton distribution in $\rho$
for  the rapidity $\eta$ has the
 Gaussian form:
\begin{equation}
\label{12}
\varphi(\rho,\eta)=\frac{C(\eta)}{\pi\gamma(\eta_p-\eta)}e^{-\frac{\rho^2}
{\gamma(\eta_p-\eta)}}   ,
\end{equation}
if the impact parameter of the initial parton is considered as the
origin. Consequently, the partons with $\eta \approx 0$ have the
broadest distribution, and, hence, the fast hadron is of the size
\begin{equation}
\label{13}
R=\sqrt{\gamma\eta_p}\approx\sqrt{\gamma\ln\frac{2p}{m}}.
\end{equation}
The account of the recombination and the scattering of the partons
affects only densities of partons and fluctuations, but does not
change the radius of the distribution  which can be viewed
as the front of the diffusion wave.

Let us discuss the parton distribution over
the longitudinal
coordinate. A relativistic particle with a momentum $p$ is
commonly considered as
a disk of the  thickness $1/p$. In fact, this is
true only in the first
approximation of the
perturbation theory. Really, a
 hadron is a disk with radius
$\sqrt{\gamma\ln\frac{2p}{m}}$ and
the  thickness of the order of $1/\mu$.
Indeed, each parton with a longitudinal momentum $k_{iz}$ is distributed
in the longitudinal direction in an interval $\Delta z_i \sim \frac{1}
{k_{iz}}$. Since the parton spectrum exists in the range of momenta
from $p$ down to $k_i \sim \mu$,
the longitudinal projection of the hadron
wave function has the structure depicted in Fig.11.
\begin{figure}[h]
\begin{center}
\begin{picture}(250,200)
\put(50,-13){
\epsfxsize=7.0cm
\epsfbox{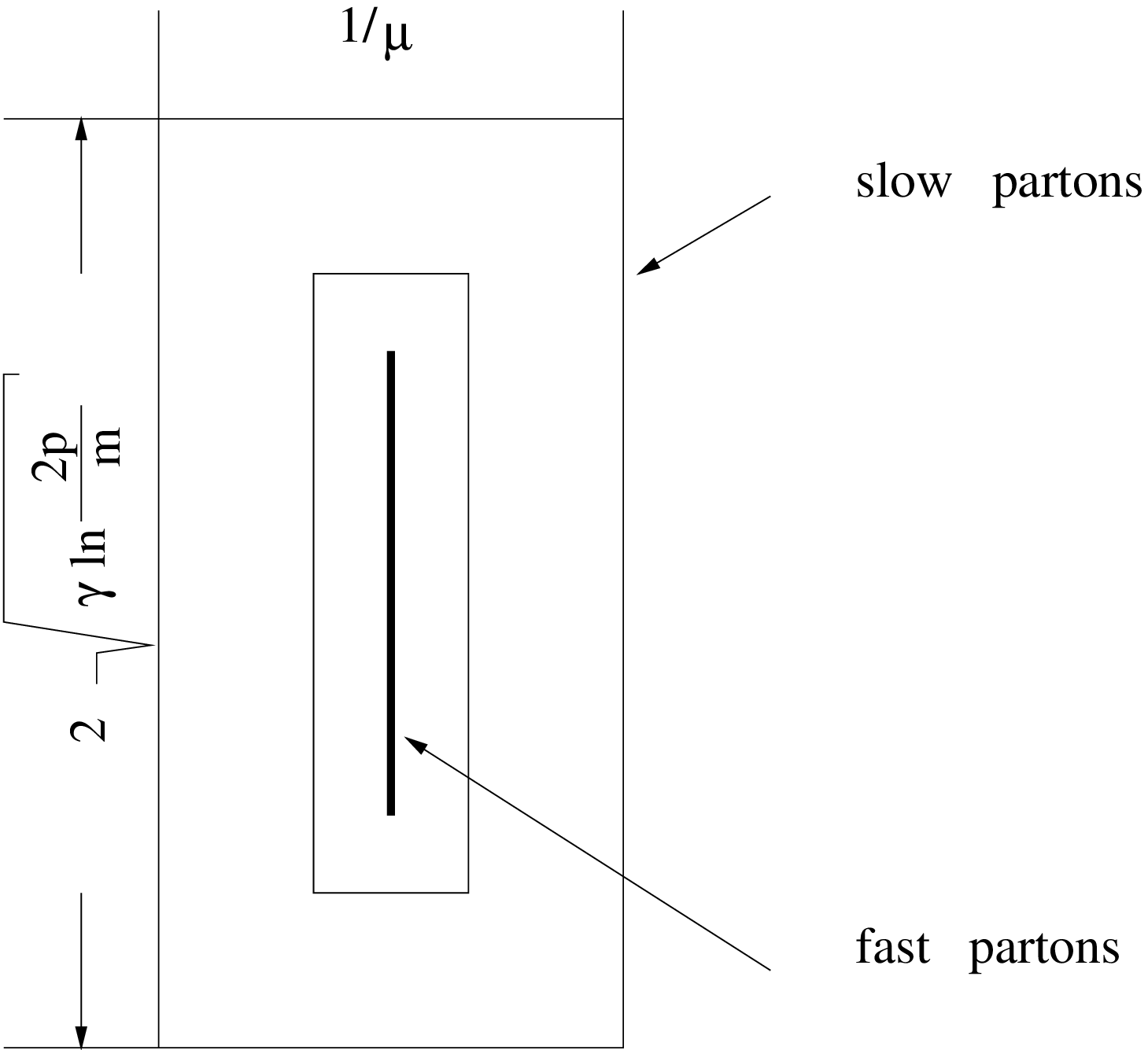}
}
\end{picture}
\end{center}
\caption{}
\end{figure}

Finally, let us consider what is the lifetime of a particular parton.
As we have discussed in the Introduction, in a theory which is not singular
at short distances, the intervals $y^2_{12}$ between two events represented
by a Feynman diagram are of the order of unity. For a fast particle moving
along the $z$ axis, $z_{21}=vt_{21}$ and $y^2_{12} = t_{21}^2 \frac{m^2}
{p^2}$. Consequently, the lifetime of a fast parton with a momentum $k_i$
is of the order of $\frac{k_i}{\mu^2}$.
The presented arguments were based on the $\lambda\varphi^3$ theory
which is the only theory which provides  a cutoff in transverse momenta.
Still, the argument should hold for other theories and for particles
with spins, if one assumes that in these theories the cutoff of transverse
momenta occures in some way. On the other
hand, the $\lambda\varphi^3$ theory cannot be considered as a
self-consistent example. Indeed, due to the absence of a vacuum state,
the series of perturbation theory
do not make sense (series with positive coefficients
are increasing as factorials).
Hence, the picture we have
presented here does not correspond {\em literally} to any particular
field theory. At the same time, it corresponds
fully  to the main
ideas of the quantum field theory and to its basic space-time relations.

\section{Deep inelastic scattering}

It is convenient to consider the deep inelastic scattering of electrons
in the frame where the time component of the virtual photon momentum is
$q_0=0$. In this reference frame the momentum of the photon is equal to
$-q_z$ ($q^2=-q_z^2$), while the momentum of the hadron is
$p_z=\omega q_z/2$ ($\omega=-2p\cdot q/q^2$). Suppose that $q_z$ is
large and $\omega\sim 1$. According to our previous considerations,  a
fast hadron can be viewed as an ensemble of partons. In this system a
photon looks as  a static field with the wavelength $\sim 1/q_z$.

The main question is, with which partons can the photon interact. We can
consider the static field of a photon as a packet with a longitudinal
size of the order of $1/q_z$. The interaction time between a hadron with
the size $1/\mu$ and such a packet is of the order of $1/\mu$. However,
due to the big difference between the parton and photon wave lengths, the
interaction with a slow parton is small. Hence, the photon interacts with
partons which have momenta  of the order of $q_z$. Partons with such
momenta are distributed in the longitudinal direction in the region
$1/q_z$. Because of this, the time of the hadron-photon interaction is
in fact of the order of $1/q_z$, i.e. much shorter than the lifetime of a
parton. This means that the photon interacts with a parton as with a free
particle, and so not only the momentum but also the energy is conserved.
As a result, the energy-momentum conservation laws select the parton
with momentum $q_z/2$, which can absorb a photon
\[  k_{iz}-q_z = k'_{iz} , \quad  |k_{iz}-q_z| = k_{iz} .\]
This gives
\[  k_{iz}=\frac{q_z}{2} , \quad  k'_{iz}=-\frac{q_z}{2}.\]
The cross section of such a process is, obviously, equal to the cross
section $\sigma_0$ of the absorption of a photon by a free particle,
multiplied by the probability to find a parton with a longitudinal
momentum $q_z/2$ inside the hadron, i.e. by the value $\varphi(\eta_{q/2},
\eta_p)$ (\ref{9}), integrated over $k_{\perp}$. (The necessary accuracy
of fulfilment of the conservation laws allows any $k_{\perp}\ll q_z$ ).

As it was already discussed, $\varphi(\eta,\eta_p)=\varphi(\eta-\eta_p)
\equiv
\varphi(\omega)$. Hence, using the known cross section for the interaction
of the photon with a charged spinless particle, we obtain for the cross
section of the deep inelastic scattering
\begin{equation}
\label{14}
\frac{d^2\sigma}{dq^2d\omega}=\frac{4\pi\alpha^2}{q^4}\left(1-\frac{pq}
{pp_e}\right)\varphi(\omega),
\end{equation}
where $p_e$ is the electron momentum. If the partons have spins, the
situation becomes more complicated,
since the cross sections of the interactions between photons and partons
with different spins are different. The parton distributions in rapidities
for different spins may also be different,
leading to the form:
\begin{equation}
\label{15}
\frac{d^2\sigma}{dq^2d\omega}=\frac{4\pi\alpha^2}{q^4}\left\{
\left(1-\frac{(pq)}{(pp_e)}\right)\varphi_0(\omega)+
\left[1-\frac{pq}{pp_e}+\frac{1}{2}\left(\frac{pq}{pp_e}\right)^2
\right]\varphi_{\frac{1}{2}}(\omega)\right\}.
\end{equation}
Let us discuss now a very important question, namely: what physical
processes take place in deep inelastic scatterings. To clarify this, we
go back to
Fig.7 determining the hadron wave function. We will neglect the parton
recombinations in the process of their creation from the initial parton,
i.e. we consider fluctuations of the type shown in Fig.9. Suppose that
the photon was absorbed by a parton with a large momentum $q_z/2$. As a
result, this parton obtained a momentum $-q_z$ and moves in the opposite
direction with momentum $-q_z/2$. The process is depicted in
Fig.12. What will now happen to this parton and to the remaining
partons? Within the framework we are using it is highly unlikely that
the parton with the momentum $-q_z/2$ will have time to interact with
the other partons. The probability to interact directly with residual
partons will be small,
because the relative momentum of the parton with $-q_z/2$ and the rest
of the partons is large. It could interact with other partons after many
subsequent decays which, in the end, could create a slow parton. However,
the time needed for these decays is large, and during this time the
parton and its decay products will move far away from the remaining
partons, thus the interaction will not take place.

Hence, we come to the conclusion that one free parton is moving in the
direction $-q_z$. What will we observe experimentally, if we investigate
particles moving in this direction? To answer this question, it is
sufficient to note that, in average, a hadron with a momentum $k_z$
consists of $n$ partons; $n=C\ln\frac{k_z}{\mu}$ at $k_z \gg \mu$.

In a sense there should exist an uncertainty relation between
the number of partons in a hadron ($n$) and the number of hadrons in a
parton ($n_p$)
\begin{equation}
\label{16}
n_pn {\stackrel{>}{\sim}}   c\ln\frac{k_z}{\mu}    ,
\end{equation}
where $k_z$ is the momentum of the state.

We came to the conclusion that the parton decays into a large number
of hadrons i.e. in fact the parton is very short-lived, highly virtual.
Hence, we have to discuss whether this conclusion is consistent with
the assumption that the photon-parton interaction
satisfies the energy conservation. To answer this question, let us
calculate the mass of a virtual parton with momentum $k_z$, decaying
into $n$ hadrons with momenta $k_i$ and masses $m_i$.
\[  M^2 = (\sum \sqrt{m_i^2+k_i^2})^2 - k_z^2 = \left(k_z + \sum_i
 \frac{m_i^2+k_{i\perp}^2}{2k_{iz}}\right)^2 - k_z^2 \approx k_z\sum_i
\frac{m^2_i+k^2_{i\perp}}{k_{iz}}.
\]
If the hadrons are distributed almost homogeneously in rapidities, their
longitudinal momenta decrease exponentially with their number, and in
the sum only a few terms, corresponding to slow hadrons, are relevant. As
a result, $M^2\sim k_z \mu$, i.e. the time of the existence of the parton
is of the order of $1/\mu$, much larger than the time of interaction with
a photon $1/q_z$.
\begin{figure}[h]
\begin{center}
\begin{picture}(250,200)
\put(-50,0){
\epsfxsize=5.0cm
\epsfbox{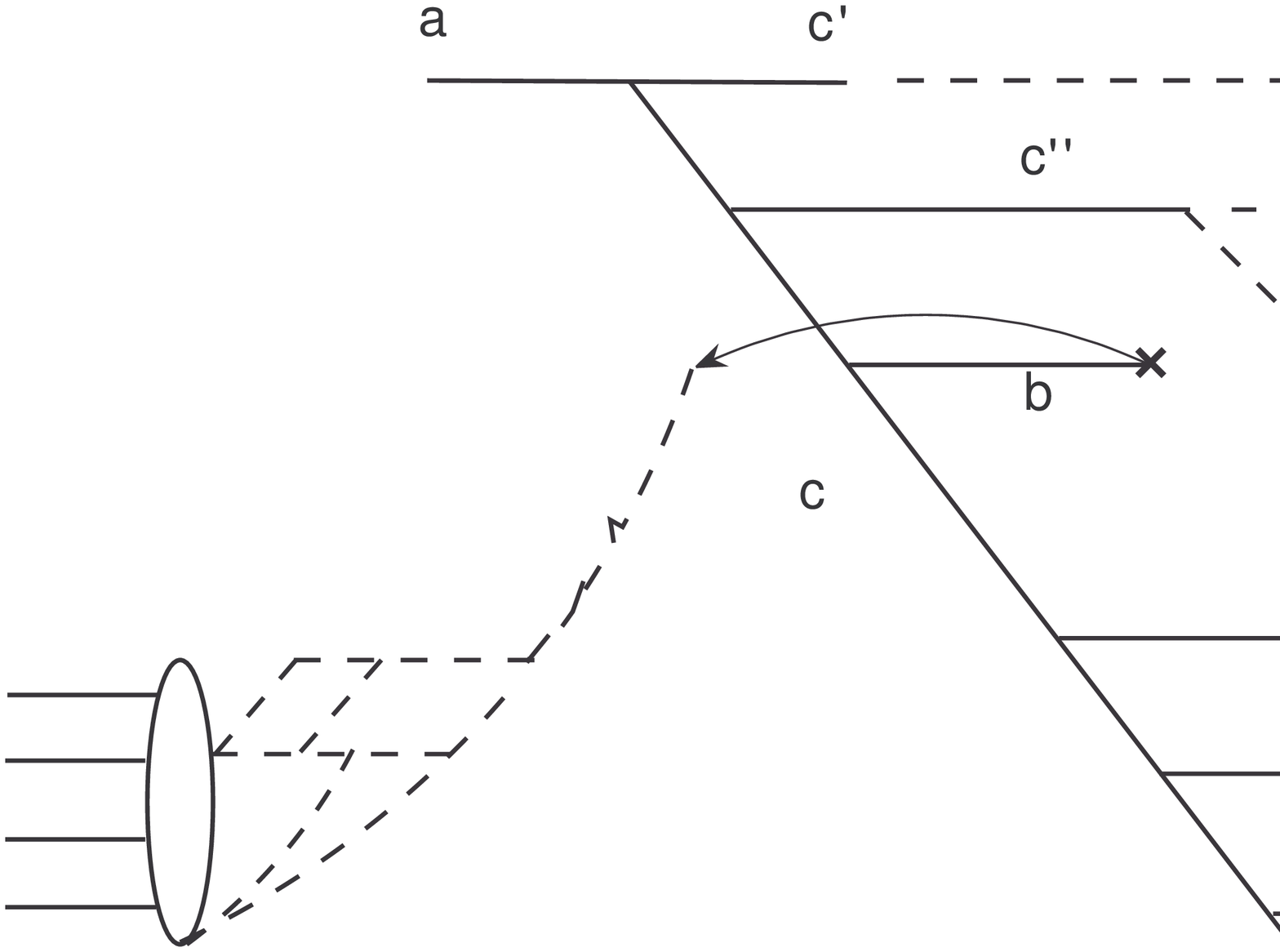}
}
\end{picture}
\end{center}
\caption{}
\end{figure}
Let us discuss now, what happens to the remaining partons.
Little can be determined
using only  the uncertainty relation eq.(\ref{16}).
This is because the number of partons before the photon absorption
was $n$, after the photon absorption it became $n-1$ and, consequently,
according to the uncertainty relation, the number of hadrons corresponding
to this state can range from 1 to $n$.
Hence, everything depends on the real perturbation of the hadron wave
function due to the photon absorption.

Consider now the fluctuation shown in Fig.12. The photon absorption
will not have any influence on partons created after the parton "$b$"
which absorbed the photon was produced, and and which
have  momenta smaller than ``b''. These fluctuations will
continue, and the partons can, in particular, recombine back into the parton
"$c$". The situation is different for partons which occured earlier and
have large momenta ("$c^{\prime}$", "$c^{\prime\prime}$"). In this case the
fluctuation cannot evolve further the same way, since the parton
"$b$" has moved in the opposite direction. As a result, it is highly probable
that partons "$c^\prime$" and "$c^{\prime\prime}$"
will  move apart and lose coherence.  On the other hand, slow partons
which were emitted by "$c^\prime$" and "$c^{\prime\prime}$"
earlier and which are not connected with the parton "$b$",
will be correlated, as before, with each of them. Thus
"$c^\prime$" and "$c^{\prime\prime}$" will move
in space together with their slow partons, i.e. in the form of hadrons.
Hence, it appears that partons flying in the initial
direction lead to the production of the order of
$c\ln\frac{\omega q_z}{2} - c\ln\frac{q_z}{2} = c\ln \omega$
hadrons with  rapidities ranging from
$\ln \frac{\omega q_z}{\mu}$ to $\ln \frac{q_z}{\mu}$.
This answer can be interpreted in the following way. After the photon
is absorbed, a hole is created in the distribution of partons
moving in the initial direction.

Contrary to the case of rapidities of partons, we will count the rapidity of
the hole not from  zero rapidity but from the rapidity
$\ln \frac{\omega q_z}{\mu}$.
In this case the rapidity of the hole is $\ln \omega$. If we now represent
the parton hole with rapidity $\ln \omega$ as a superposition of
the hadron
states, this superposition will contain $\ln \omega$ hadron states.

Let us represent the whole process by a diagram describing
rapidity distributions of
partons and hadrons. Before the photon absorption the partons
in the hadrons are distributed at rapidities
between zero and $\ln \frac{\omega q_z}{\mu}$, while after the
photon absorption a parton distribution is produced which is shown in
Fig. 13.

This parton distribution leads to the hadron distribution shown in
Fig. 14. The total multiplicity corresponding to this distribution is
\[ \bar{n} = c\ln\frac{q_z}{\mu} + c\ln\omega = c\ln\frac{\nu}{\mu\sqrt
 {-q^2}} . \]
\begin{figure}[h]
\begin{center}
\begin{picture}(250,160)
\put(-60,45){
\epsfxsize=7.0cm
\epsfbox{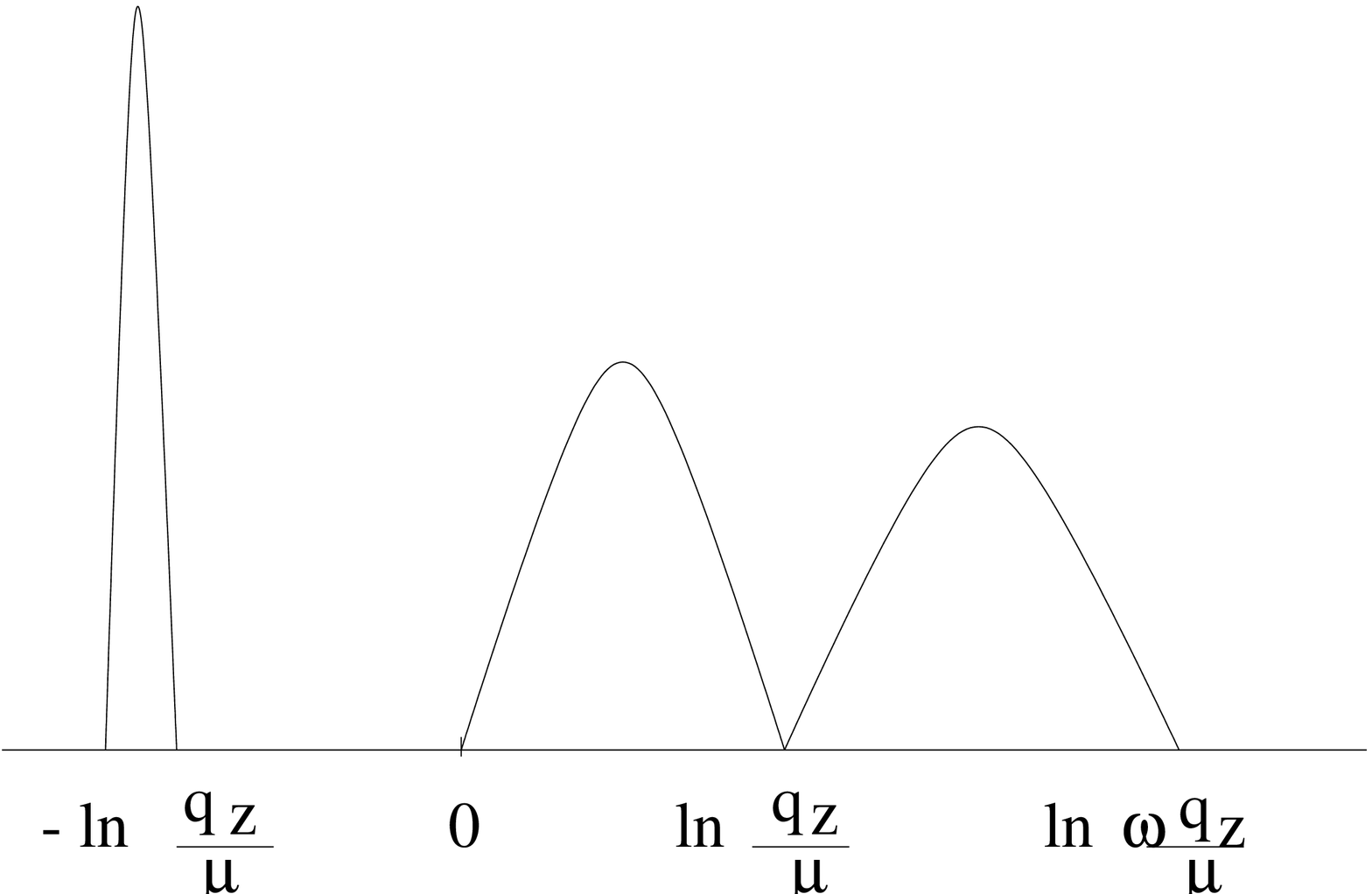}
}
\put(155,45){
\epsfxsize=7.0cm
\epsfbox{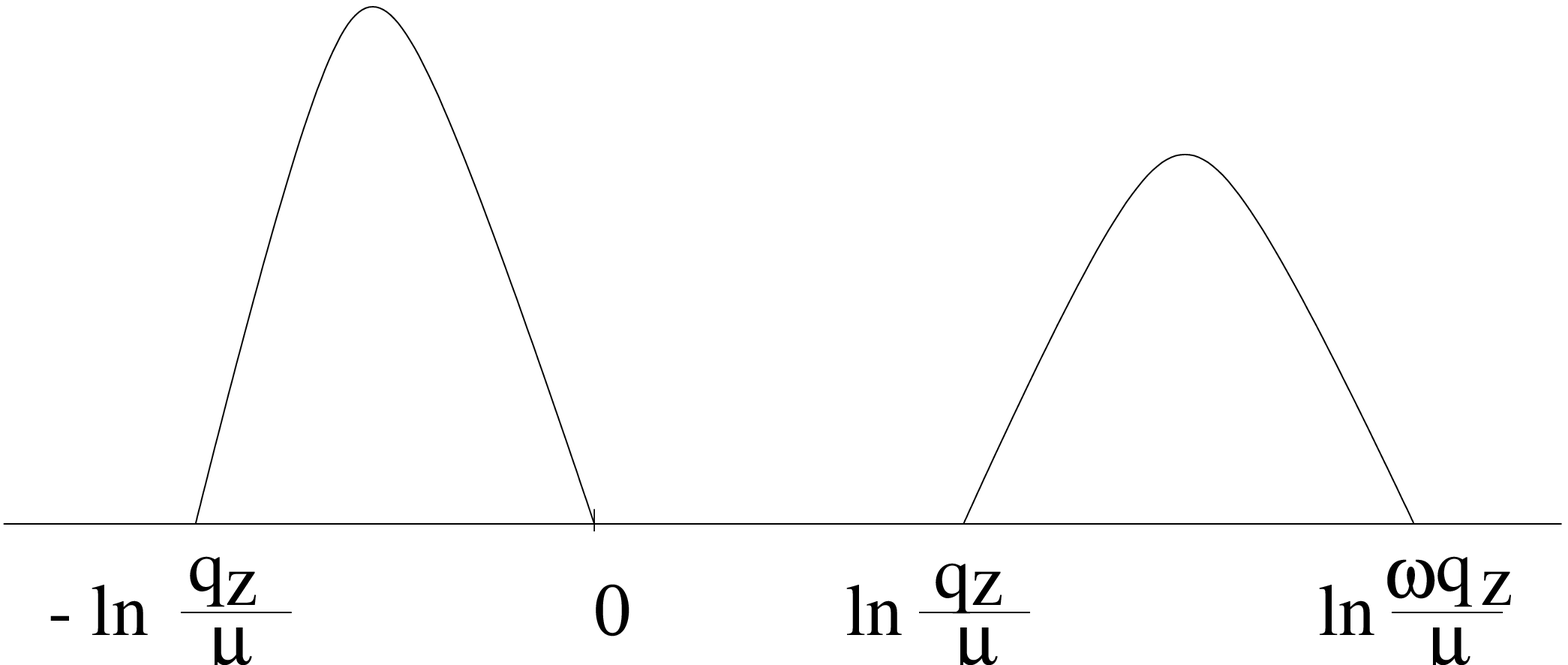}
}
\put(12,10){Figure 13 \hspace{4 cm} Figure 14}
\end{picture}
\end{center}
\end{figure}
\setcounter{figure}{14}

This hadron distribution in rapidities in the deep inelastic scattering
 differs qualitatively from those previously discussed in the
literature. It corresponds to $c\ln\frac{\sqrt{-q^2}}{\mu}$
hadrons moving in the photon momentum direction, and  $\ln \omega$
hadrons are moving in the nucleon momentum
direction,
with a gap in rapidity between  these distributions. The hadron
distribution which was obtained in the framework of perturbation theory for
superconverging theories like $\lambda\varphi^3$ (Drell, Yan)
differs qualitatively from the distribution in Fig. 14.

In  conclusion of this part, it is necessary to point out
that the problem of spin properties of the partons exist in
this picture even if the partons do
not have quark quantum numbers. If, as the experiment shows, the cross
section $\sigma_T$ for the interaction of the transversal photons is
larger than the cross section for the interaction of the
longitudinal photons, $\sigma_L$, the charged partons have predominantly
spin $1/2$. This means that at least one fermion, for example a
 nucleon, has to move in the direction
of the photon momentum. In other words, in deep inelastic scattering
the distribution of the
created hadrons in quantum numbers as the function
of their rapidities differs essentially from what we are used to in the
strong interactions. Perhaps this is one of the key prediction of the
non-quark parton picture for  $\sigma_t \gg \sigma_l$.

\section{Strong interactions of hadrons}

Let us discuss now the strong interactions of hadrons.
First, we consider a collision of two hadrons in the laboratory frame.
Suppose that a hadron "1" with momentum $\vec{p}_1$ hits hadron
"2" which is at rest. Obviously, the parton wave function makes no
sense for the hadron at rest, since for the latter the vacuum fluctuations
are absolutely essential. However, the hadron at rest can also be
understood as an ensemble of slow partons distributed in a volume
of the order of $1/\mu$, independent of the origin of the partons.
Indeed, it does not matter whether these partons are  decay products
of the initial parton or the result of the vacuum fluctuations. How can a fast
hadron, consisting of partons with rapidities from $\ln\frac{2p_1}{\mu}$
to zero, interact with the target which consists of slow partons?
Obviously, the cross section of the interaction of two point-like particles
with a large relative energy is not larger than $\pi\lambda^2 \sim
\frac{1}{s_{12}} \sim e^{-\eta_{12}}$ (where $\lambda$ is the wave length
in the c.m. frame, $\eta_{12}$ is the relative rapidity). That is why
only  slow partons of the incident hadron can interact with the target
with a cross section which is not too small. This process is shown in
Fig. 15.
\begin{figure}[h]
\begin{center}
\begin{picture}(250,150)
\put(0,0){
\epsfxsize=7.0cm
\epsfbox{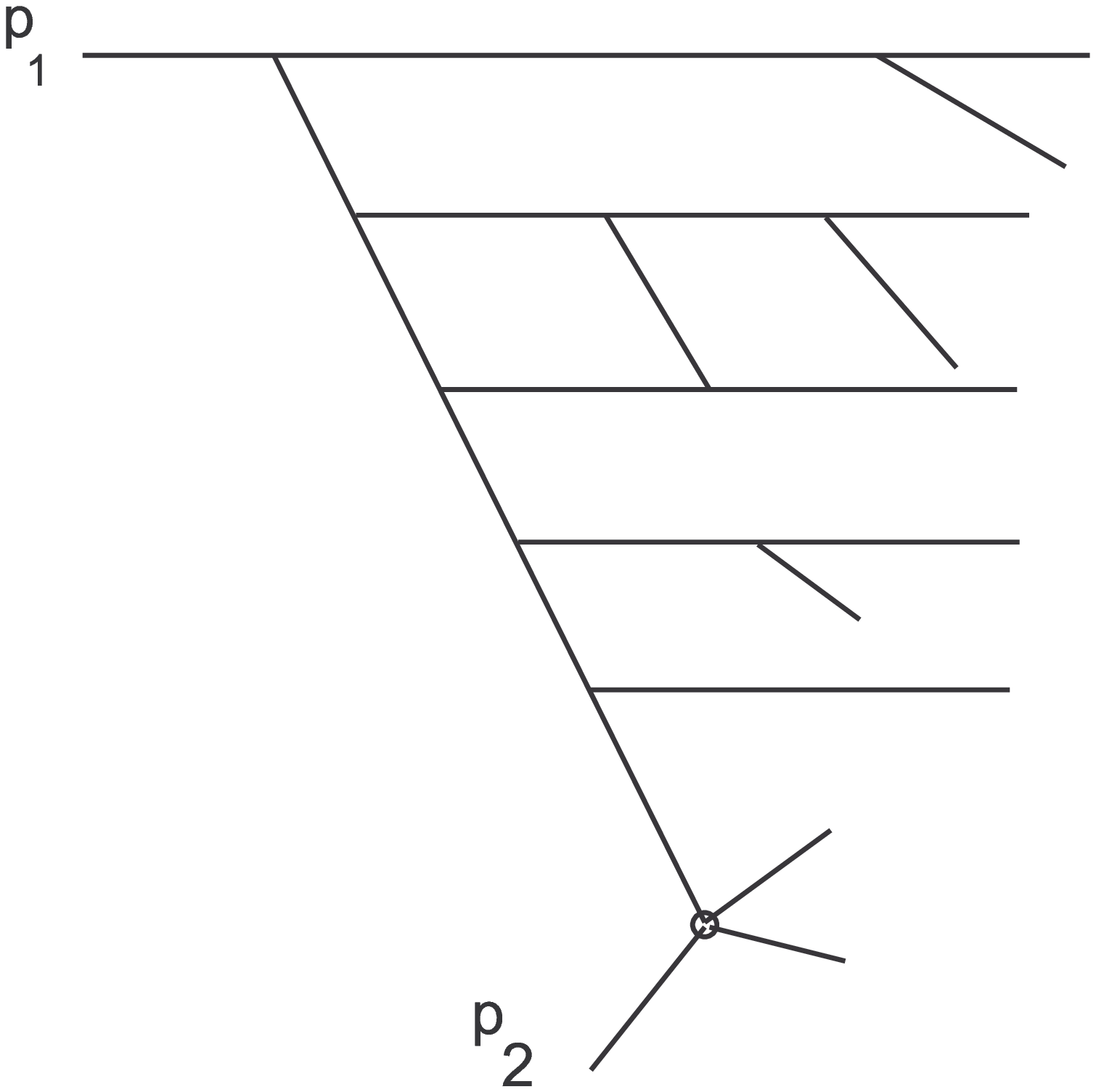}
}
\end{picture}
\end{center}
\caption{}
\end{figure}

If the slow parton which initiated the interaction was absorbed in
this interaction, the fluctuation which lead to its creation from a fast
parton was interrupted. Hence, all partons which were emitted by the
fast parton in
the process of fluctuation cannot recombine any more. They disperse
in space and ultimately decay into hadrons leading to the creation of
hadrons with rapidities from zero to $\ln\frac{2p_1}{\mu}$. The
interaction between the partons is short-range in rapidities.
Hence, the hadron distribution in rapidities will reproduce the parton
distribution in rapidities.
In particular, the inclusive spectrum of
hadrons will have the form shown in Fig. 8, with an unknown
distribution near the boundaries. The total hadron multiplicity will be of
the order of $\eta_p = \ln \frac{2p_1}{\mu}$. If the probability of
finding a slow parton in the hadron does not depend on the
hadron momentum (this would be quite natural, since with the increase of
the momentum the life-time of the fluctuation is also growing), the total
cross section of the interaction will not depend on the energy at high
energies.

Before continuing the analysis of inelastic processes, let us
discuss, how to reconcile the energy independence of the total
interaction cross section  at high energies with the
observation discussed above that the transverse hadron sizes increase
with the increase of
the energy as $\sqrt{\gamma\ln\frac{2p}{\mu}}$. The answer is that
slow partons are distributed almost homogeneously over the
disk of the radius $\sqrt{\gamma\ln\frac{2p}{\mu}}$ (Eq.(\ref{11})) ,
while their overall multiplicity during the time of $1/\mu$
is of order of unity.

Let us see now how the
same process will look, for example, in the c.m. frame. In this
reference frame the interaction will have the form as shown in Fig. 16.
\begin{figure}[h]
\begin{center}
\begin{picture}(250,150)
\put(-115,45){
\epsfxsize=7.0cm
\epsfbox{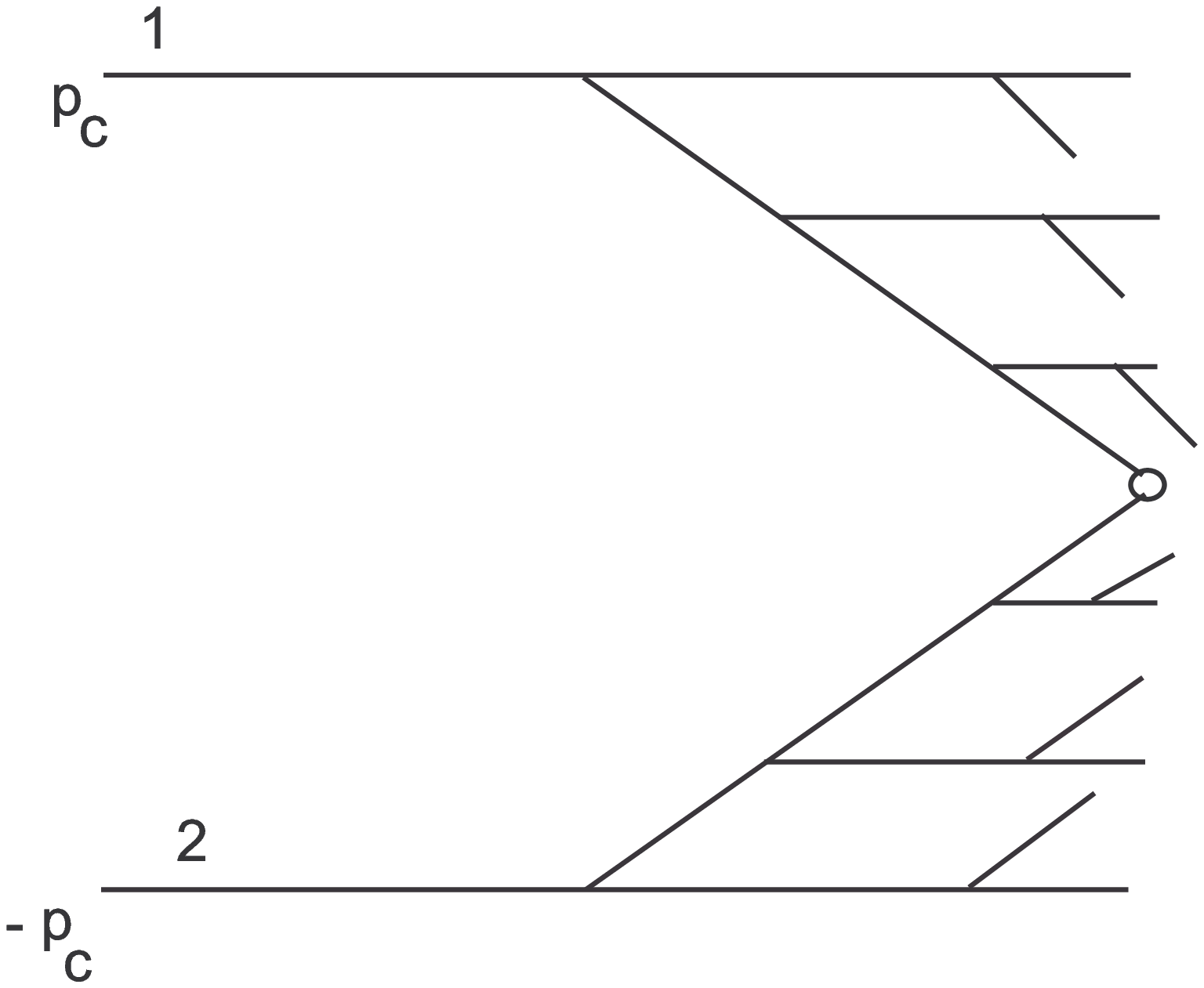}
}
\put(115,45){
\epsfxsize=7.0cm
\epsfbox{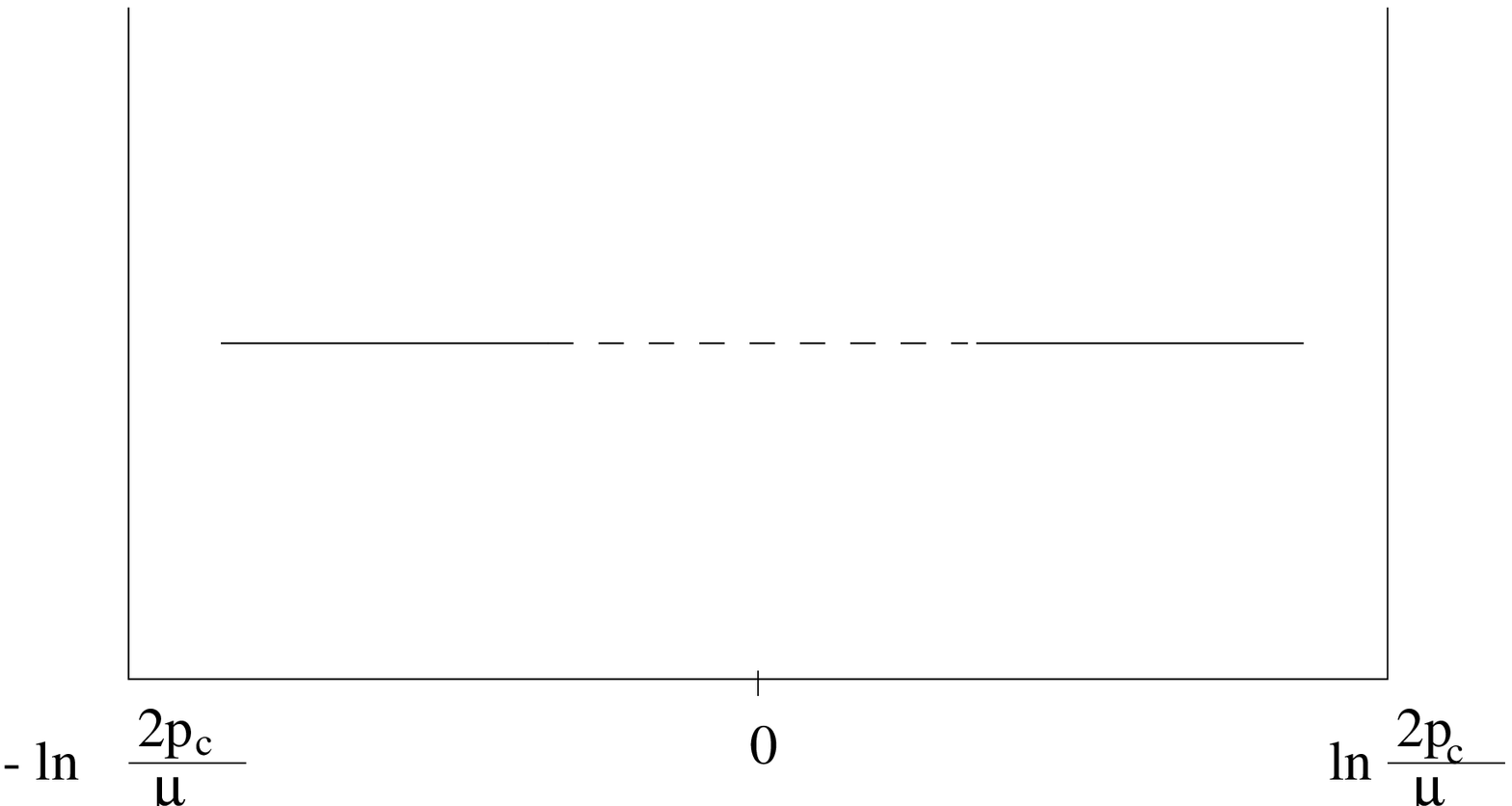}
}
\put(-10,12){
Figure 16. \hspace{5 cm} Figure 17}
\end{picture}
\end{center}
\end{figure}
\setcounter{figure}{17}

Each of the hadrons consists of partons with rapidities ranging from
$-\ln\frac{2p_c}{\mu}$ to zero and from zero to $\ln\frac{2p_c}{\mu}$,
respectively. The slow partons interact with cross sections
which are not small.
As a result, the fluctuations will be interrupted in both hadrons, and
the partons will fly away in the
opposite directions, leading to the creation of
hadrons with rapidities from $-\ln\frac{2p_c}{\mu}$ to $\ln\frac{2p_c}
{\mu}$. From the point of view of this reference
frame the inclusive spectrum
must have the form shown in Fig. 17, with unknown distributions not only
at the boundaries but also in the centre, since the distribution of the
slow partons in the hadrons and in vacuum fluctuations is unknown. The
hadron inclusive spectrum, however, should not depend on the
reference frame.
Thus the inclusive spectrum in Fig. 17 should
coincide with the inclusive spectrum in Fig. 8, and they should differ
only by a trivial shift along the rapidity axis, i.e. due to relativistic
invariance we know something about the spectra of slow partons and
vacuum fluctuations.

Let us demonstrate that this comparison of processes in two reference
frames leads to a very important statement, namely that at ultra-high energies
the total cross sections for the interactions of arbitrary
hadrons should be equal. Indeed, we have assumed that the distribution
of hadrons reproduces the parton distribution.

From the point of view of the
laboratory frame the distribution of partons and, consequently, distribution of
hadrons in the central region of the spectrum is completely determined by
the properties (quantum numbers, mass, etc.) of particle 1,
and does not depend on the properties of particle 2.
On the other hand,
from the point of view of the
antilaboratory frame (where the particle 1 is at rest)
 everything is determined by the properties of particle
2. This is possible only if the distribution of partons
in the hadrons with rapidities $\eta$ much smaller than the hadron rapidity
$\eta_p$ does not depend on the quantum numbers and the mass of the
hadron, that is  the parton distribution with $\eta \ll \eta_p$ should be
universal. From the point of view of the c.m. system the same region is
determined by slow partons of both hadrons and by vacuum fluctuations
(which are universal), and, consequently, the distribution of slow
partons is also universal.

It is natural to assume
that the probability of finding a hadron in a sterile state without slow partons
tends to zero with the increase of its momentum, in other words assume
 that slow partons are always present in a hadron
(compare to the decrease
of the cross section of the elastic electron scattering at large $q^2$).
In this case considering the process in the
c.m. system, we see that  the total cross section
of the hadron interaction is determined by the cross section of the
interaction of slow partons and by their transverse
distribution which is universal.
Consequently, the total hadron interaction cross section is also
universal, i.e. equal for any hadrons.

This statement looks rather strange if we regard it, for instance, from
the following point of view. Let us consider the scattering of a
complicated system with a large radius, for example, deuteron-nucleon
scattering. As we know, the cross section of the deuteron-nucleon
interaction equals the sum of the nucleon-nucleon cross sections,
thus  it is
twice as large as the nucleon-nucleon cross section. How and at what
energies can the deuteron-nucleon cross section become equal to the
nucleon-nucleon cross section? How is it possible that the density of
slow partons in the deuteron
turns out to be equal to the density of slow partons in the
nucleon? To answer this question, let us discuss the parton structure
of two hadrons which
are separated in the plane transverse to their longitudinal momenta
by a distance much larger
than their Compton wave length $1/\mu$.
Suppose that at the initial moment they were point-like particles.
Next, independently of each other, they begin to emit
partons with decreasing longitudinal momenta. At the same time the
diffusion takes place in the transverse plane so that the
partons will be distributed in a growing region. The basic observation
which we shall prove and which answers our question is that if the
momenta of the initial partons are sufficiently large, then during one
fluctuation the partons coming from different initial partons will
inevitably meet in space (Fig. 18) in the region of the order of $1/\mu$.
They will have similar large rapidities and, hence, will be able to
interact with a probability of the order of unity. If such ``meetings''
take place sufficiently frequently, the probability of
the parton interaction
will be unity. Consequently, the further evolution
 and the density of
the slow partons which are created after the
meeting may not depend on the
fact that initially the transverse distance between
 two partons was large.

\begin{figure}[h]
\begin{center}
\begin{picture}(250,260)
\put(-100,20){
\epsfxsize=6.0cm
\epsfbox{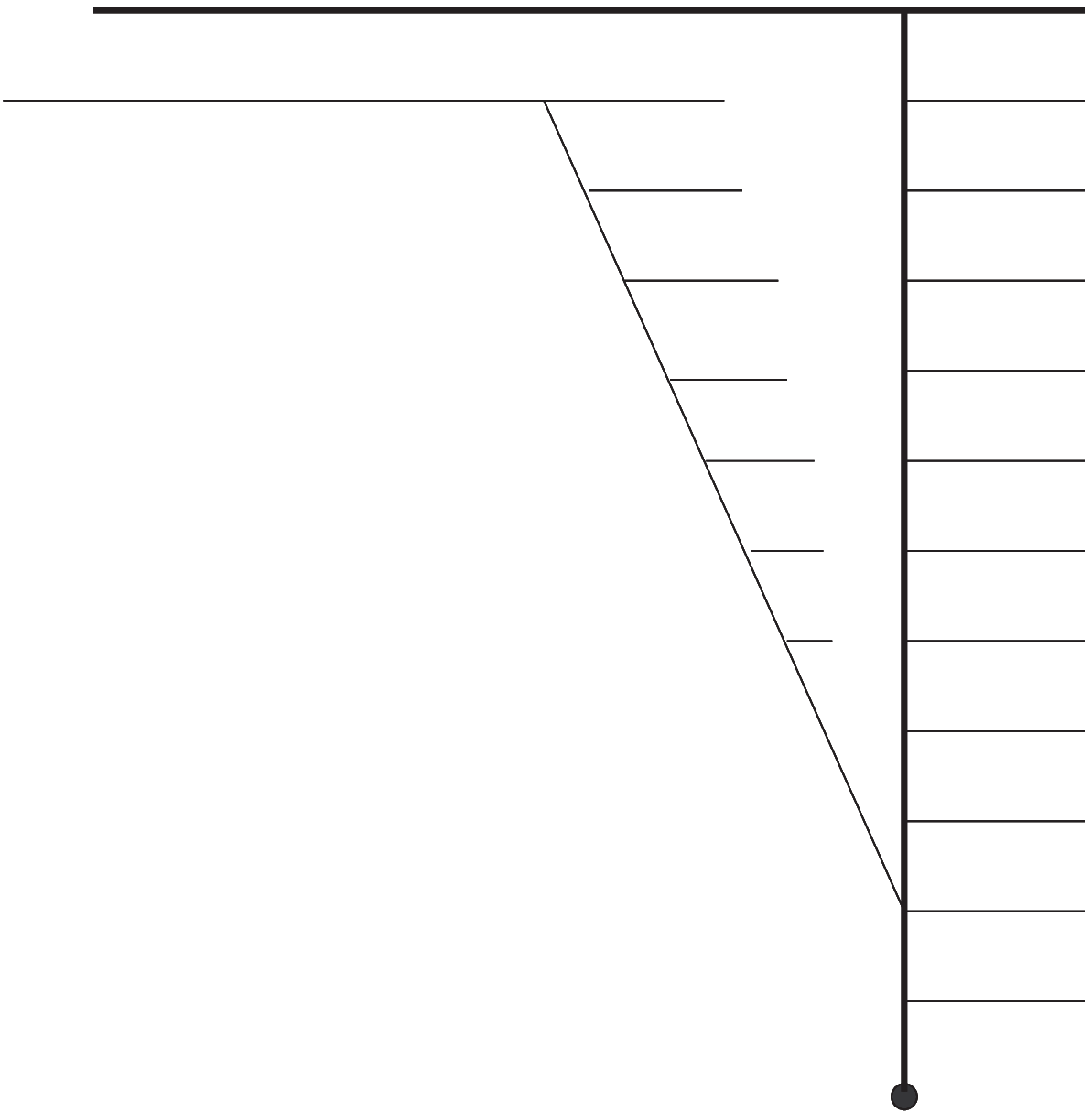}
}
\put(145,33){
\epsfxsize=6.0cm
\epsfbox{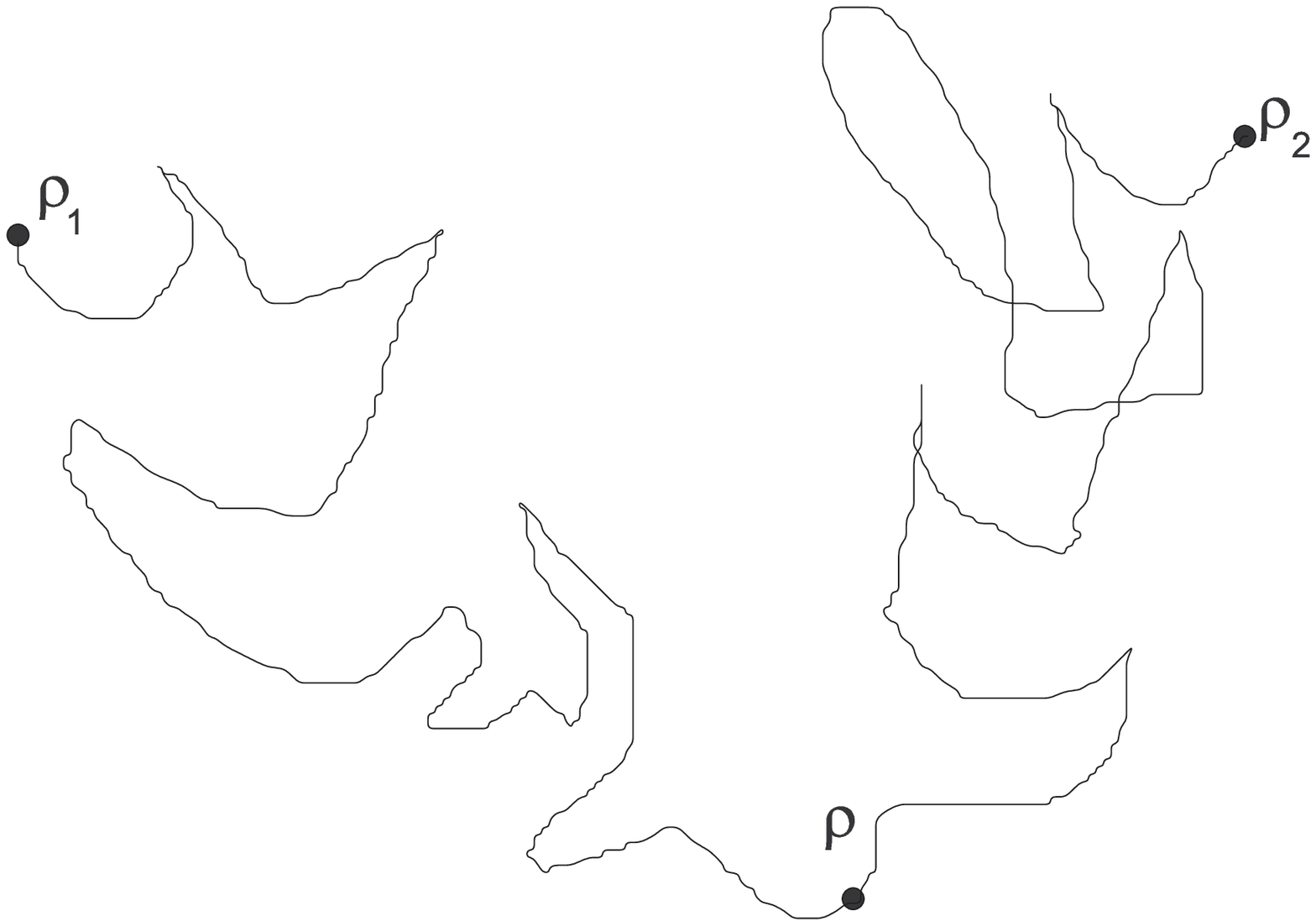}
}
\put(0,-15){
Figure 18 \hspace{5.0 cm} Figure 19}
\end{picture}
\end{center}
\end{figure}
\setcounter{figure}{19}

In terms of the diffusion in the impact parameter
plane this statement corresponds to the following picture.
Suppose that initial partons were placed
at points $\rho_1$ and $\rho_2$ in Fig. 19 and that their
longitudinal momenta are of the same
order of magnitude, i.e. difference of their rapidities is of the
order of unity, while each of the rapidities is large.
We will follow the parton starting from point $\rho_1$,
which decelerates
via emission of other partons. As we have seen, its
propagation in the perpendicular plane corresponds to  diffusion.
The difference of rapidities $\eta_p-\eta$ at the initial and the
considered moments  serves the role of time in this
diffusion process.

The diffusion character of the process means that the probability density
of finding a parton with rapidity $\eta$ at the point $\rho$ if it started
from the point $\rho_1$ with
rapidity $\eta_p$ is
\begin{equation}
\label{17}
\omega(\vec{\rho},\vec{\rho}_1,\eta_p-\eta)=
\frac{1}{\pi\gamma(\eta_p-\eta)}e^{
-\frac{(\vec{\rho}-\vec{\rho}_1))^2}{\gamma(\eta_p-\eta)}} .
\end{equation}
The situation is exactly the same for a decelerating parton which started
from the point $\rho_2$. Thus, the probability of finding both partons at
the same point $\rho$ with equal rapidities is proportional to
\begin{eqnarray}
\label{18}
\lefteqn{ \omega(\rho_{12},\eta_p-\eta) = } \nonumber\\
& & \int \omega(\vec{\rho},\vec{\rho}_1,\eta_p-\eta)
\omega(\vec{\rho},\vec{\rho}_2,\eta_p-\eta) d^2\rho= \nonumber\\
& & \frac{1}{2\pi\gamma(\eta_p-\eta)}\exp\left[
-\frac{(\vec{\rho}_1-\vec{\rho}_2))^2}{2\gamma(\eta_p-\eta)}\right]
\end{eqnarray}
If we now integrate this expression over $\eta$, i.e. estimate the
probability for the partons to meet
at some rapidities, we obtain
\begin{equation}
\label{19}
\int_{0}^{\eta_{p}}\omega(\rho_{12},\eta_p-\eta)d\eta\approx\frac{1}{\pi}
\log\frac{2\gamma\eta_p}{\rho^2_{12}}|_{\eta_p\rightarrow\infty}
\longrightarrow \infty  .
\end{equation}
This means that if $2\gamma\eta_p \gg \rho^2_{12}$, the partons
will inevitably
meet. According to (\ref{19}) we get a probability much larger than unity.
The reason is that under these conditions the meeting of partons
at different values of $\eta$ are not independent events and therefore
it does not make sense to add the probabilities. It is easy to prove this
statement directly, for example with the help of the diffusion equation.
We will not do this, however.
According to a nice analogy suggested
by A. Larkin, this theorem is equivalent to the statement that if you are in
an infinite forest in which there is a house on a finite distance from
you, then, randomly wandering in the forest, you sooner or later arrive
at this house. Essentially, the reason is that in the two-dimensional
space the region inside of which the diffusion takes place and the length
of the path travelled during the diffusion increase with time in the same
way.
>From the point of view of the reference frame in which the deuteron is
at rest and is hit by a nucleon in the form of a disk, the radius of
which is much larger than that of the deuteron, the statement of the
equality of cross sections means that the parton states inside the disk
are highly coherent.

It is clear from above that the cross sections of two
hadrons can become equal only when the radius of parton distribution
$\sqrt{\gamma\eta_p}$ which is increasing with the energy becomes much
larger than the size of both hadrons. Substituting
$4\cdot\frac{0,3}{m^2}$ for the value of $\gamma$
($m$ is the proton
mass)
\footnote{It will be demonstrated below that $\gamma=4\alpha^\prime$, 
where $\alpha^\prime$ is the slope of the Pomeron trajectory. The
current data give
$\alpha^\prime \sim 0.25 GeV^{-2}$.
}
we see that the deuteron-nucleon
cross section will practically never coincide with the nucleon-nucleon
cross section, while the tendency for convergence of
cross sections for pion-nucleon, kaon-nucleon and
nucleon-nucleon scatterings may be manifested already
starting at the incident energies $\sim
10^3$ GeV.

\section{Elastic and quasi-elastic processes}

So far we focused
on the implications of the considered picture for
inelastic processes with multiplicities, growing logarithmically with
the energy.
However, with a certain probability it can
happen that slow partons scatter at very small angles and the
fluctuations will not be interrupted in either of the hadrons (for example,
if we discuss the process in the 
c.m. frame). In this case small angle
elastic or quasi-elastic scattering will  take place (Fig. 20).
\begin{figure}[h]
\begin{center}
\begin{picture}(250,200)
\put(-20,-13){
\epsfysize=7.0cm
\epsfbox{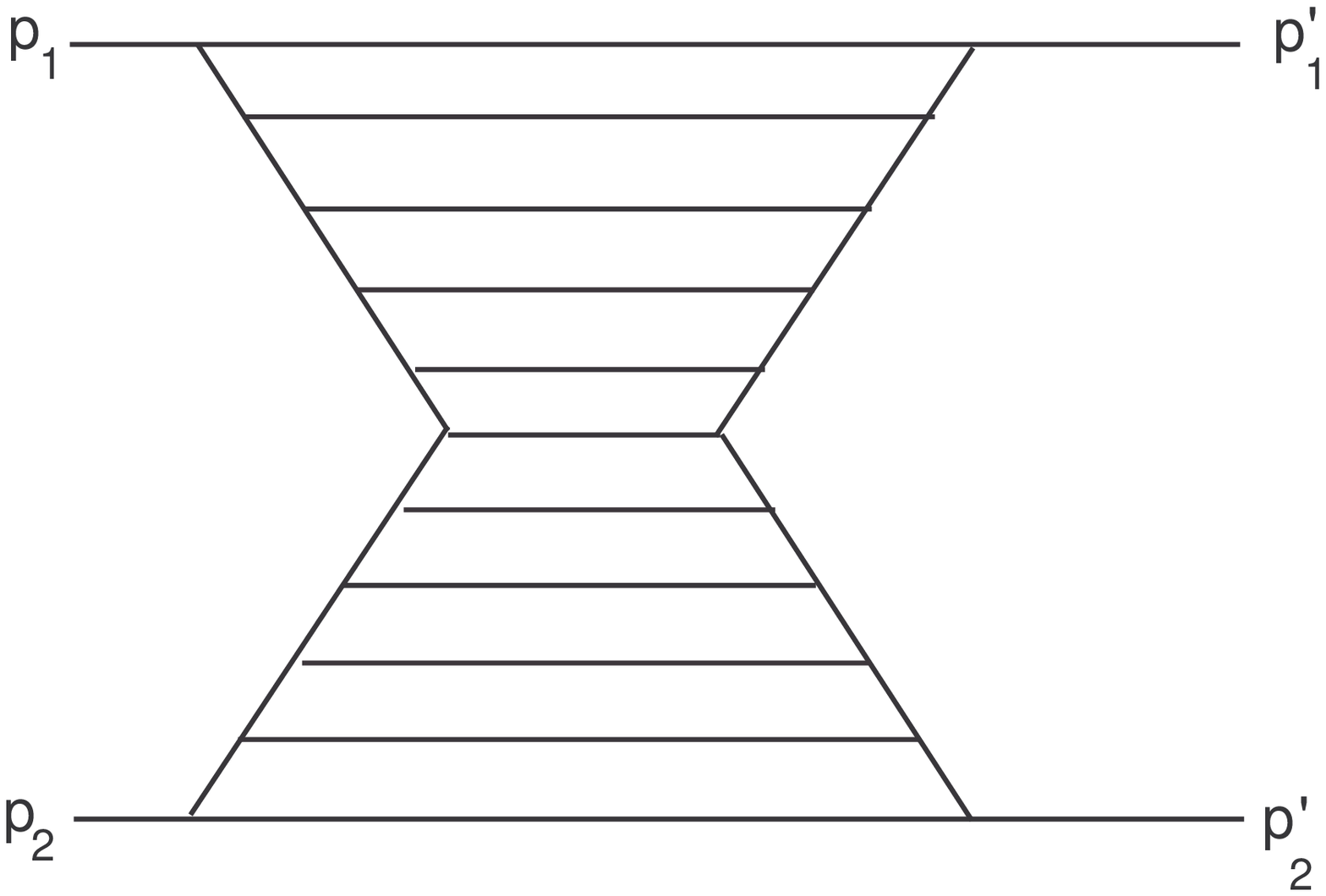}
}
\end{picture}
\end{center}
\caption{}
\end{figure}
First, let us calculate  the elastic scattering amplitude. It is well
known that the imaginary part of the elastic scattering amplitude can be
written in the form
\begin{equation}
\label{20}
A_1(s_{12})=s_{12}\int d^2\rho_{12}e^{i\vec{q}\vec{\rho}_12}
\sigma(\rho_{12},s_{12}),
\end{equation}
where $s_{12}$ is the energy squared in the c.m. system, $\rho_{12}$
is the relative impact parameter, $\sigma(\rho_{12},s_{12})d^2\rho_{12}$
- the total interaction cross section of particles being at the distance
$\rho_{12}$ and $\vec{q}$ is the momentum transferred.
In order to calculate
$\sigma(\rho_{12},s_{12})$ it is sufficient to notice that, according
to (\ref{12}), the probability of finding a slow parton with
rapidity $\eta$ at the impact parameter $\rho'_1$ which originated
from the first hadron with an impact parameter $\vec{\rho}_1$ is
\begin{equation}
\label{21}
\varphi_1(\vec{\rho}_1,{\vec{\rho^\prime}_1},\eta_1,\eta_{pc} )
\frac{C(\eta_1)}{\pi\gamma\eta_{pc}}
\exp\left[
-\frac{(\vec{\rho}_1-{\vec{\rho^\prime}_1})^2}{\gamma\eta_{pc}}\right] .
\end{equation}
The probaility a parton originating from the second hadron
at impact parameter $\rho'_2$ is
\begin{equation}
\label{22}
\varphi_2(\vec{\rho}_2,\vec{\rho}_2^\prime,\eta_2,\eta_{pc} )
\frac{C(\eta_2)}{\pi\gamma\eta_{pc}} \exp\left[
-\frac{(\vec{\rho}_2-\vec{\rho}^\prime_2)^2}{\gamma\eta_{pc}}\right] .
\end{equation}

The total cross section of the hadron interaction which is due to
the interaction of slow partons is equal to
\[ \sigma(\rho_{12},s_{12})=\int
d\eta_1d\eta_2 d^2\rho^\prime_{12} C(\eta_1)C(\eta_2) \]
\[ \times \int\frac{d^2\rho}{(\pi\gamma\eta_{pc})^2}
\exp\left[
-\frac{(\vec{\rho}-\vec{\rho}_1)^2}{\gamma\eta_{pc}}
-\frac{(\vec{\rho}-\vec{\rho}_2)^2}{\gamma\eta_{pc}}
\right].
\]
We have taken into account that $\rho'_1=\rho+\frac{\rho'_{12}}{2}$,
$\rho'_2=\rho-\frac{\rho'_{12}}{2}$,
and that the dependence on $\rho'_{12}$ can be neglected
in the exponential factor.

After carrying out the integration over $\rho$, we obtain
\begin{equation}
\label{23}
\sigma(\rho_{12},s_{12})=\frac{e^{-\frac{(\vec{\rho}_1-\vec{\rho}_2)^2}
{2\gamma\eta_{pc}}}\sigma_0}{2\pi\gamma\eta_{pc}}
\end{equation}
Inserting (\ref{22}) into (\ref{20}), we get
\[ A_1=s_{12}\sigma_0 e^{-\frac{\gamma}{4}q^2\xi} ,\]
\begin{equation}
\label{24}
\xi=2\eta_{pc}=\log\frac{s_{12}}{\mu^2} .
\end{equation}
We obtained the scattering amplitude corresponding to the exchange by
the Pomeranchuk pole with the slope $\alpha'= \gamma/4$, $\sigma_0=g^2$
where $g$ is the universal coupling constant of the
Pomeron and hadron.
The amplitude (\ref{24}) is usually represented by diagram in Fig. 21
\setcounter{figure}{20}
\begin{figure}[h]
\begin{center}
\begin{picture}(100,50)
\put(-160,-50){
\epsfysize=4.0cm
\epsfbox{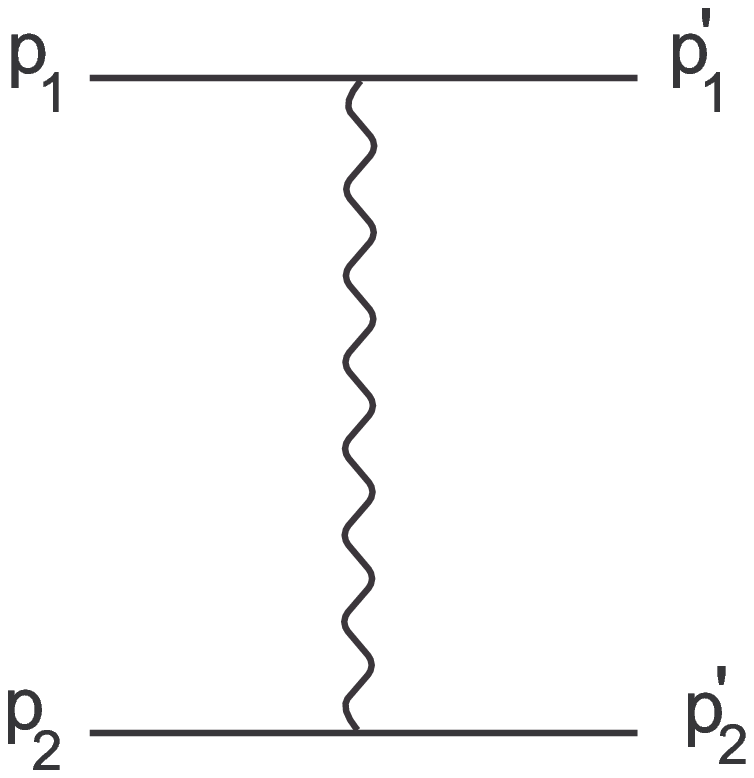}
}
\end{picture}
\end{center}
\caption{}
\end{figure}

where a propagator of the form $e^{-\alpha' q^2 \xi}$ corresponds to the
Pomeron. In the impact parameter space this propagator has the form
(\ref{22}).

Let us discuss the physical
meaning of $\sigma_0$ in more detail. For this purpose, let us
calculate the zero angle scattering amplitude at ($\vec{q}=0$),
without using the impact parameter representation.
The probability of finding a parton
with rapidity $\eta$ and a transverse momentum $k_{\perp}$ is described
by (\ref{9}). This expression at $\eta \ll \eta_p$
corresponds to the diagram in Fig.22. The wavy line represents integration
over parton rapidities from $\eta_p$ to zero.

\begin{figure}[h]
\begin{center}
\begin{picture}(100,140)
\put(-160,-10){
\epsfxsize=14.0cm
\epsfysize=6.0cm
\epsfbox{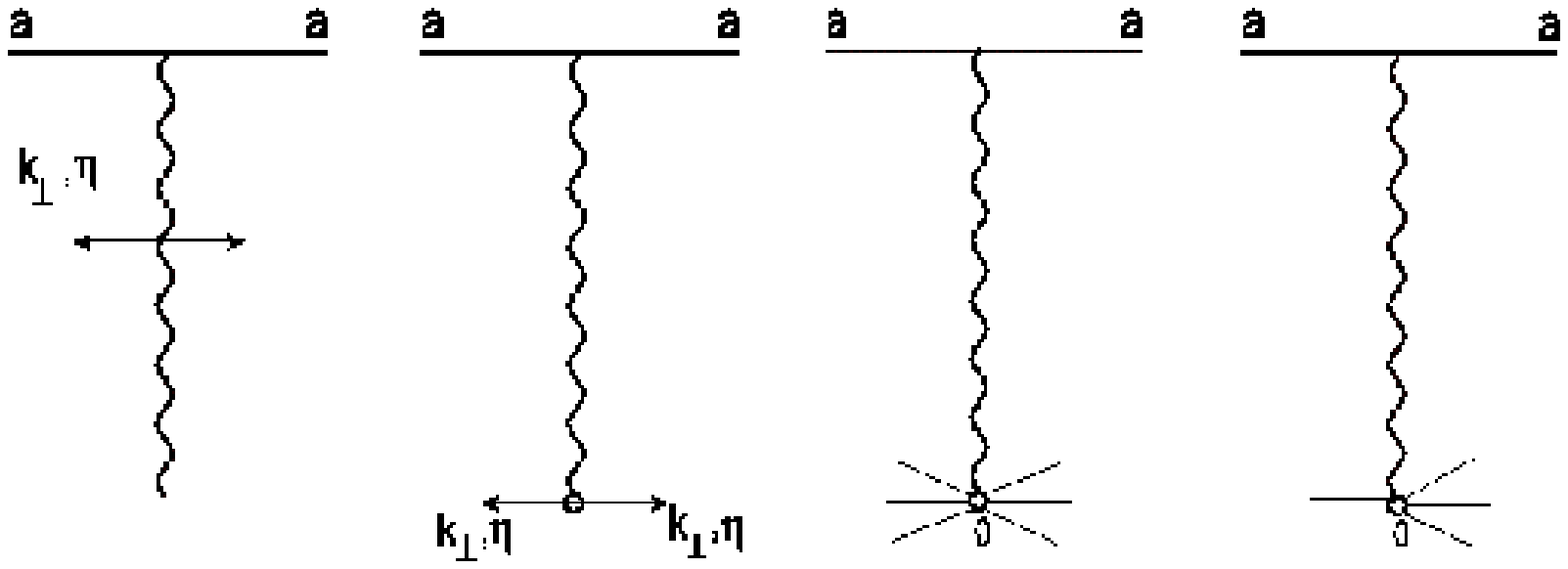}
}
\put(-145,-15){
Figure 22 \hspace{1.8 cm} Figure 22a \hspace{2.0 cm} Figure 22b
\hspace{1.8 cm}Figure 22c
}
\end{picture}
\end{center}
\end{figure}
This figure reflects the hypothesis that  the calculation of
 $\varphi(\eta,k_t,\eta_p)$ for suffiently large $\eta_p$ and $\eta \ll
\eta_p$
leads to an expression for
$\varphi$ which is factorized in the same way as the
Pomeron contribution to the scattering amplitude.
This is because the parton
 distribution in this  region is independent of
 the properties of the hadron as well as
the values of $\eta,\eta_p$.
Compared to the diagram in Fig. 7, Fig. 22
indicates that the calculation of $\varphi(\eta,k_t,\eta_p)$ is similar to
 the calculation of the inclusive cross section due to the Pomeron
exchange.
The only difference is that the coupling of the hadron with the Pomeron
should be substituted by unity, since a hadron always exists in a Pomeron
state.
If  $\eta\sim 1$,  $\varphi(\eta,k_t)$ corresponds to the diagram in
Fig.22a,
 which shows that $\varphi(\eta,k_t)$ depends on $\eta$. Similarly, it is
possible
 to determine the probability of finding several slow partons (Fig. 22b),
 and even the density matrix of slow partons. In this case the amplitude
 of elastic hadron-hadron scattering in the center of mass frame is
determined
by the diagram of fig.23 and the value of $\sigma_0$ is determined solely
by the interaction of slow partons.
\begin{figure}[h]
\begin{center}
\begin{picture}(100,150)
\put(-180,-60){
\epsfysize=7.0cm
\epsfbox{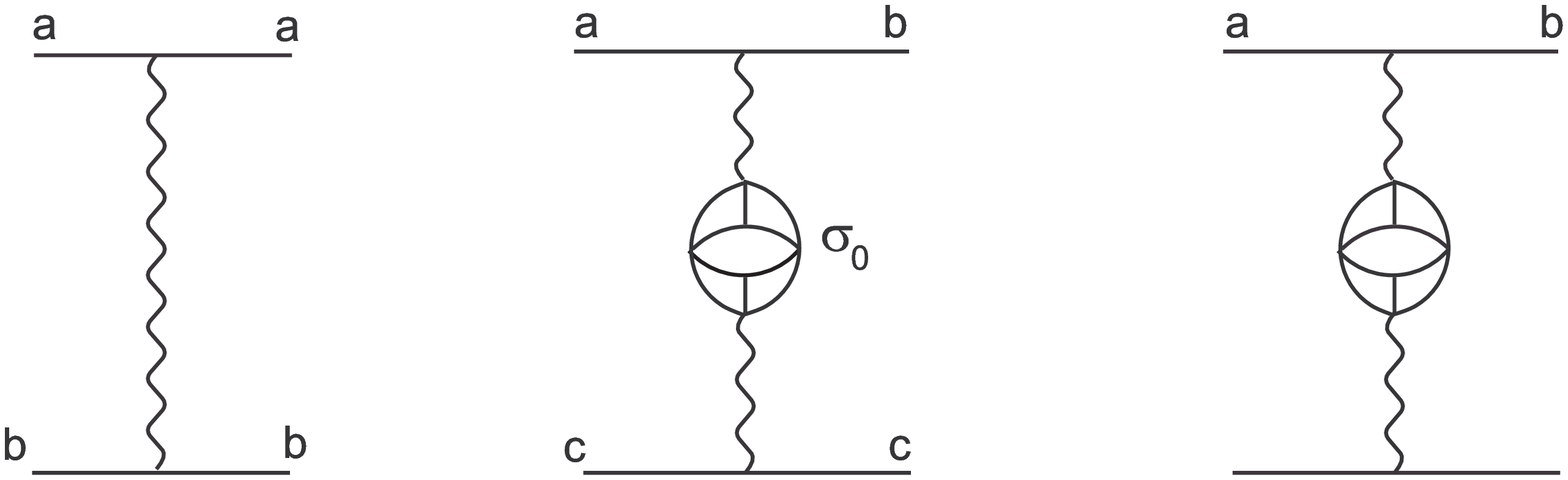}
}
\put(-130,-15){
Figure 23 \hspace{3.0 cm} Figure 24 \hspace{3.5 cm} Figure 25}
\end{picture}
\end{center}
\end{figure}

Now let us consider  the quasi-elastic scattering, corresponding to the Pomeron
exchange (Figs.24,25) at zero transverse momentum. While the probability
to find the parton in hadron "a" is determined in eq.(8) by the integral
of the wave function squared, the analogous quantity for the amplitude of the
inelastic diffractive process (Fig.25) will lead to the integral of the product of
the parton functions of different hadrons. They are orthogonal to each
other and it is almost obvious that amplitude for inelastic diffractive
process
at zero angle should vanish for this reason. Indeed the orthogonality
condition of eq.(6) has the same structure as the imaginary part of
the amplitude. Thus, if at high energies the amplitude factorizes
(as it should be for the
Pomeron exchange), than the orthonormality condition
should also have factorized form in the sense that the integral
over parton rapidities with $\eta\ll \eta_p$  factors out,
 and only constants
$g_{ab}$ depend on the properties of specific hadrons (see Fig.26).
\setcounter{figure}{25}
\begin{figure}[h]
\begin{center}
\begin{picture}(100,100)
\put(-20,-13){
\epsfysize=4.5cm
\epsfbox{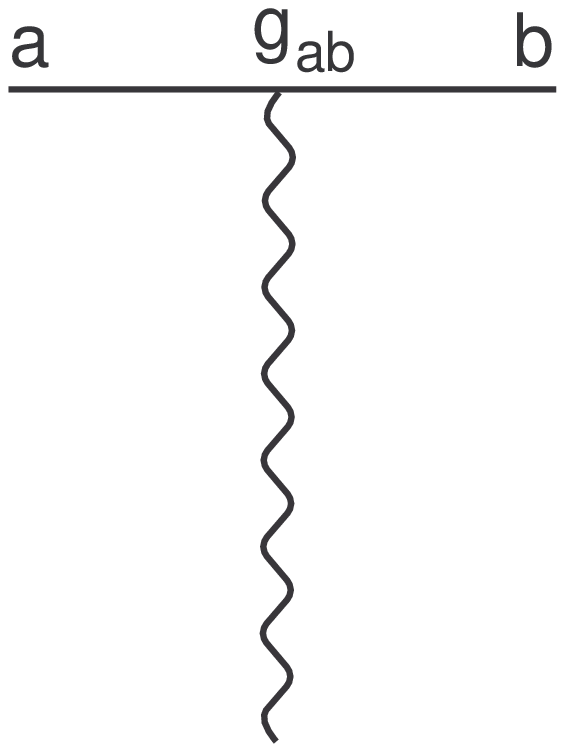}
}
\end{picture}
\end{center}
\caption{}
\end{figure}
Orthogonality of the wave functions of different hadrons implies that
$g_{ab}=0$ at $a\neq b$. In fact the reason, why the amplitude of
inelastic diffractive process vanishes at zero angle
is the same as the reason why all cross sections should
approach the same value at high energies. Both phenomena are due
to the fact that properties of slow partons do not depend on the
properties of hadrons to which they belong. We can illustrate
this again using the example of quasi-elastic dissociation
of the composite system --- e.g. deuteron. Let us consider the
interaction of a fast nucleon with a deuteron. As we discussed in
the previous section, at
very large energies partons from different nucleons will
always interact with each other  independent of the distance between
nucleons. This will lead to production of the spectrum of slow
partons which does not depend on the relative distance between
nucleons in deuteron.
This means that the amplitude of the nucleon-nucleon interaction will
not depend on the internucleon distance as well. Thus, if nucleons
inside the
deuteron will remain intact  after the interaction,
than the deuteron will not dissociate as well, since
if the  amplitude does not depend
on the internucleon distance, the wave function of the deuteron
will not change after the interaction.


\begin{thebibliography}{99}
\bibitem{Feynm}
 R.Feynman, {\em What neutrinos can tell about partons},  Neutrino-72, v.II,
 p.75, Proceedings Conference, Hungary, June 1972. .
\bibitem{Bjorken}
J.D.Bjorken, Phys.Rev., {\bf 179} (1969), p.1547.
\bibitem{Drell}
S.D.Drell and T.M.Yan, Annals of Phys., {\bf 66}(1971), p.555
\bibitem{Gribov}
V.N.Gribov, Proceedings of Batavia Conference, 1972.
\end{thebibliography}
\end{document}